\documentclass[a4paper,11pt]{article}
\pdfoutput=1

\usepackage{jheppub,amsfonts,mathrsfs,caption,subcaption}

\newcommand{\op}{\mathcal{O}}
\newcommand{\pa}{\partial}
\newcommand{\al}{\alpha}\newcommand{\ga}{\gamma}\newcommand{\ep}{\epsilon}\newcommand{\te}{\theta}\newcommand{\om}{\omega}
\newcommand{\Ga}{\Gamma}\newcommand{\De}{\Delta}\newcommand{\Om}{\Omega}

\numberwithin{equation}{section}

\title{Kinematic space for conical defects}

\author[a]{Jesse C. Cresswell}
\author[a,b]{Amanda W. Peet}

\affiliation[a]{Department of Physics, University of Toronto, \\ 60 St. George St., Toronto, Canada}
\affiliation[b]{Department of Mathematics, University of Toronto,\\ 40 St. George St., Toronto, Canada}

\emailAdd{jcresswe@physics.utoronto.ca}
\emailAdd{awpeet@physics.utoronto.ca}

\abstract{Kinematic space can be used as an intermediate step in the AdS/CFT dictionary and lends itself naturally to the description of diffeomorphism invariant quantities. From the bulk it has been defined as the space of boundary anchored geodesics, and from the boundary as the space of pairs of CFT points. When the bulk is not globally AdS$_3$ the appearance of non-minimal geodesics leads to ambiguities in these definitions. In this work conical defect spacetimes are considered as an example where non-minimal geodesics are common. From the bulk it is found that the conical defect kinematic space can be obtained from the AdS$_3$ kinematic space by the same quotient under which one obtains the defect from AdS$_3$. The resulting kinematic space is one of many equivalent fundamental regions. From the boundary the conical defect kinematic space can be determined by breaking up OPE blocks into contributions from individual bulk geodesics. A duality is established between partial OPE blocks and bulk fields integrated over individual geodesics, minimal or non-minimal.}


\arxivnumber{1708.09838}

\begin{document}
\maketitle
\flushbottom

\captionsetup[figure]{font=small,labelfont=small}

\section{Introduction}

Even before the AdS/CFT correspondence \cite{Maldacena1999,Witten1998,Gubser1998} provided a physical duality between conformal field theories and theories of quantum gravity in Anti-de Sitter spacetimes, CFT quantities had been mathematically represented in terms of bulk fields \cite{Ferrara1971,Ferrara1972}. These ideas relating contributions to conformal blocks and integrals of bulk fields over geodesics have reemerged recently in the context of geodesic Witten diagrams \cite{Hijano2015,Hijano2016}. Whereas a four-point Witten diagram with bulk vertices integrated over the entire bulk calculates a full CFT four-point function, integrating the vertices only over geodesics connecting boundary insertions computes a conformal partial wave. The conformal partial wave represents the contribution of a primary operator and its descendants to the four-point function, and somehow knows about the geodesic structure of AdS.

 A new approach to the AdS/CFT correspondence has shed more light on the connection between composite operators in the operator product expansion (OPE), and integrated bulk fields. The authors of \cite{Czech2015,Czech2016} proposed the use of an auxiliary space that interpolates between the bulk and boundary theories, similar to the space used in \cite{Boer2015}. The auxiliary space, called kinematic space, functions as a way of organizing the non-local degrees of freedom which lead to diffeomorphism invariant quantities in the bulk gravity theory. Whereas local bulk fields fail to satisfy diffeomorphism invariance, a field integrated over a boundary anchored geodesic or otherwise attached to the boundary with a geodesic dressing can be invariant \cite{Donnelly2015,Donnelly2016}. Boundary anchored geodesics in asymptotically AdS spacetimes meet the boundary at pairs of spacelike or null separated points suggesting a relation to bi-local CFT operators. Such composite operators are easily described in terms of the OPE. Both a geodesic integrated field and the basis of non-local operators forming the OPE can be viewed as fields on kinematic space leading to a diffeomorphism invariant entry into the AdS/CFT dictionary.

Several proposals have been made as to how kinematic space should be defined from the bulk and boundary. Kinematic space was originally presented as the space of bulk geodesics with a measure derived from their lengths in terms of the Crofton form \cite{Czech2015}. Since the length of a minimal geodesic is holographically related to entanglement entropy in AdS$_3$/CFT$_2$ \cite{Ryu2006}, a boundary description of kinematic space was given as the space of boundary intervals with the metric defined in terms of the differential entropy of those intervals \cite{Balasubramanian2014}.\footnote{This approach was recently inverted to derive the universal parts of the entanglement entropy in a CFT with a boundary from knowledge of the kinematic space \cite{Bhowmick2017}.} In order to generalize the kinematic space approach to higher dimensional systems, later approaches defined points in kinematic space as oriented bulk geodesics, and simultaneously as ordered pairs of boundary points \cite{Czech2016}. 

In the case of a pure AdS$_3$ geometry, these approaches are consistent since there is a unique geodesic connecting each pair of spacelike separated boundary points. Other well known locally AdS$_3$ geometries can have several geodesics connecting each pair of boundary points, namely conical defects and the BTZ black holes \cite{Banados1992,Banados1993,Deser1984}. There are two diverging ways to modify the definition of what constitutes a kinematic space point in such cases. Any spacelike separated pair of boundary points will be connected by a unique minimal geodesic, so the bulk definition can exclude non-minimal geodesics from kinematic space with no need to change the boundary definition. Alternatively, non-minimal geodesics can be considered as points with the same standing as minimal ones, in which case ordered pairs of boundary points alone will not fill out kinematic space. Excluding non-minimal geodesics is not desirable due to the generic fact that minimal geodesics do not reach all depths of the bulk. The region probed by non-minimal geodesics is known as the entanglement shadow \cite{Balasubramanian2015,Freivogel2014,Balasubramanian2016}. A full description of the bulk in terms of kinematic space can only succeed when non-minimal geodesics are included. This forces a change to the definition of kinematic space from the boundary point of view.

In this paper, we take up the issue of non-minimal geodesics in kinematic space, and the matter of an equivalent boundary definition of points in the simplest geometry exhibiting this feature, the static conical defects in three bulk dimensions. In section \ref{secbulk} the geometry of the conical defect kinematic space is derived in two ways. The first is a simple application of the differential entropy definition applied to geodesics of all lengths. The second follows \cite{Zhang2017} in noting that the conical defects can be obtained as a quotient of pure AdS$_3$. Under this quotient classes of geodesics are identified, producing a quotient on kinematic space, and leading to a result equivalent to the first approach. In section \ref{secboundary} the metric of kinematic space is extracted from OPE blocks in the CFT. By mapping to a convenient covering CFT system we find that conventional OPE blocks can be broken down further than done before using the method of images. Individual image contributions to the OPE blocks contain information about subregions of kinematic space that, when combined, reproduce the same space identified from the bulk. Intuition from previous uses of the method of images to calculate correlation functions holographically suggests an association between partial OPE blocks in the CFT and geodesics of a fixed winding number in the bulk. Kinematic space provides a realm where the connection between these objects can be made precise, as is shown in section \ref{secdis}. We conclude by isolating the contribution to the full OPE block from individual bulk geodesics, minimal or non-minimal, connecting the boundary insertion points. This extends the holographic dictionary established in \cite{Czech2016} between OPE blocks and geodesic integrated operators, and provides more fine-grained information about the holographic contributions to the blocks.

To help visualize the physics on the conical defect, covering space, and kinematic space, an interactive Mathematica demonstration is provided with this paper as an ancillary file. It does not require a Mathematica license to be used.

\section{Kinematic space from the bulk}\label{secbulk}

In this section we focus on static conical defect spacetimes and consider the kinematic space for a constant time slice. We show that the differential entropy approach \cite{Czech2015}, and the quotient approach \cite{Zhang2017} produce different fundamental regions of the same kinematic space, but are entirely equivalent. 

\subsection{Review of geometries}

 In global coordinates, the universal cover of AdS$_3$ has the metric
\begin{equation}\label{eqgads}
ds^2=R^2_{\mathrm{AdS}}(-\cosh^2\rho\ dt^2+d\rho^2+\sinh^2\rho\ d\phi^2),
\end{equation}
with $t\in\mathbb{R},~\rho\in \mathbb{R^+}$, and $\phi\in[0,2\pi]$ with the identification $\phi=\phi+2\pi$. Throughout this paper the ``unwrapped" time coordinate $t$ of the universal cover will be used. The AdS$_3$ geometry can be understood as a surface embedded in the higher dimensional flat space $\mathbb{R}^{(2,2)}$ with metric
\begin{equation}
ds^2=-dU^2-dV^2+dX^2+dY^2.
\end{equation}
The AdS$_3$ metric is induced by restricting to a hyperbolic surface
\begin{equation}
-U^2-V^2+X^2+Y^2=-R^2_{\mathrm{AdS}}.
\end{equation}
The parameter $R_{\mathrm{AdS}}$ is the AdS length scale which will be set to unity throughout the remainder of this paper. The metric in global coordinates is obtained from the embedding equations
\begin{equation}
U=\cosh\rho\cos{t},~~V=\cosh{\rho}\sin{t},~~X=\sinh\rho\cos\phi,~~Y=\sinh\rho\sin\phi.
\end{equation}

\begin{figure}
    \centering
    \begin{subfigure}[b]{0.33\textwidth}
        \includegraphics[width=0.9\textwidth]{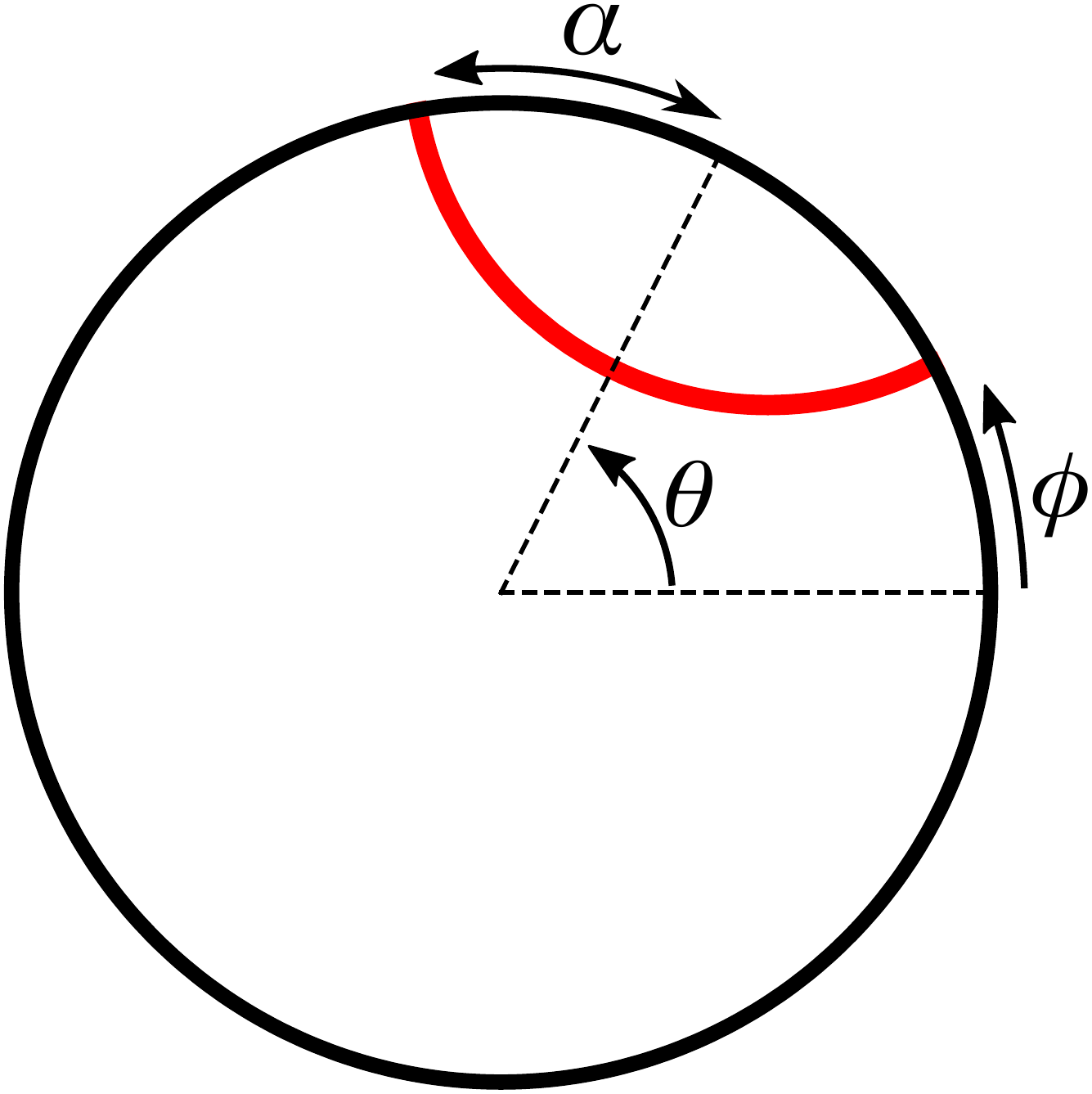}
        \caption{}
        \label{fig:H2coords}
    \end{subfigure}
     \ \ 
    \begin{subfigure}[b]{0.31\textwidth}
        \includegraphics[width=0.9\textwidth]{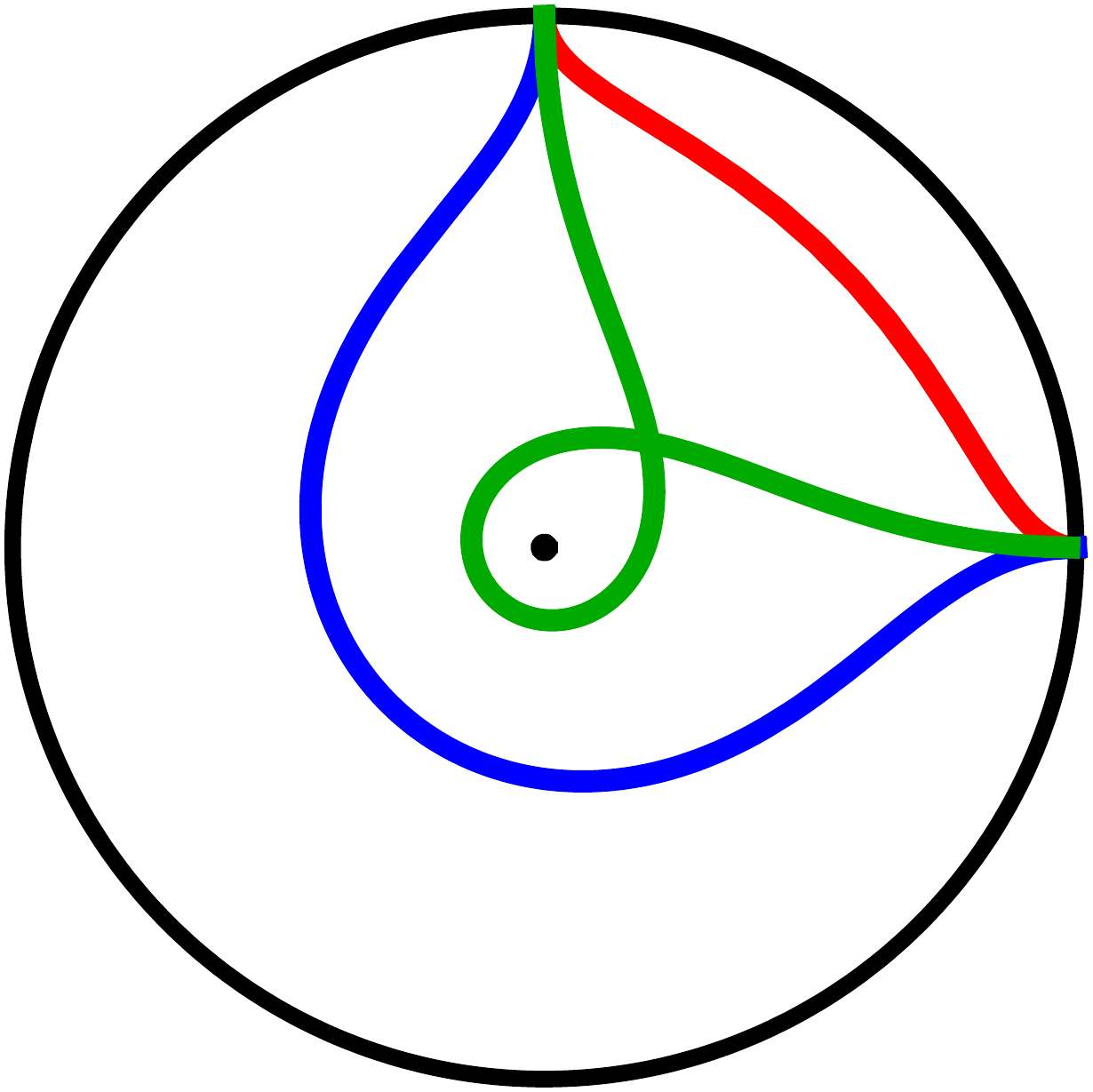}
        \caption{}
        \label{fig:CDPoincareDiskGeodesics2}
    \end{subfigure}
     \ \ 
    \begin{subfigure}[b]{0.31\textwidth}
        \includegraphics[width=0.9\textwidth]{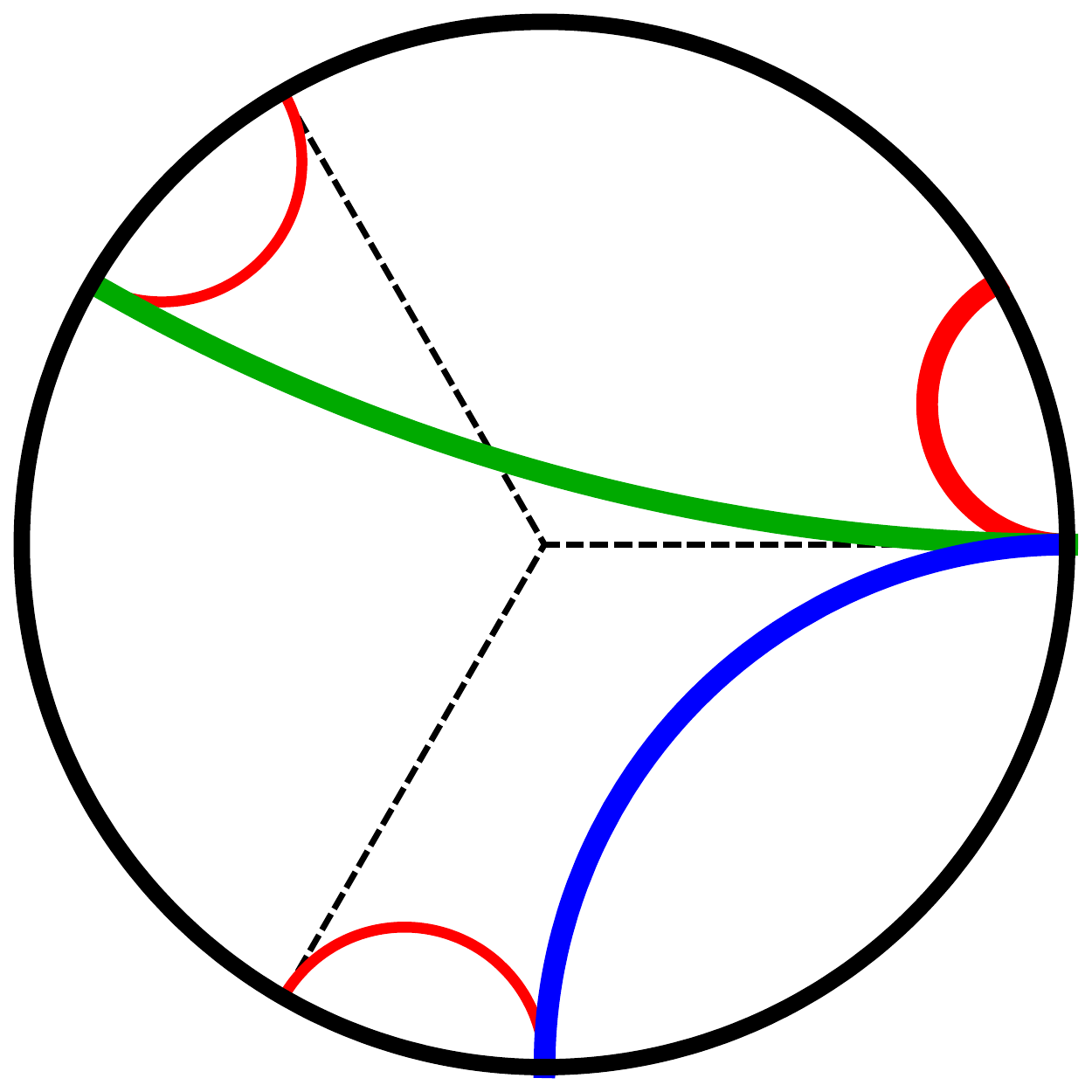}
        \caption{}
        \label{fig:PoincareDiskGeodesics}
    \end{subfigure}
    \caption{(a) The Poincare disk showing a geodesic and its kinematic coordinates. (b) A spatial slice of a conical defect, with $N=3$ for illustration, showing three geodesics subtending the same boundary interval with winding numbers $n=0,1,2$ respectively in order of increasing length. (c)~The covering space of the conical defect showing identified wedges, and preimages of the corresponding geodesics in figure (b). Also shown are two equivalent images of the minimal geodesic (red).}\label{fig:1}
\end{figure}

For visual representations it will be useful to consider the Poincar\'e disk. By taking a constant time slice $t=0$, equivalently $V=0$, the metric induced from $\mathbb{R}^{(1,2)}$ is that of the hyperbolic plane $\mathbb{H}_2$,
\begin{equation}
ds^2=d\rho^2+\sinh^2\rho \ d\phi^2.
\end{equation}
This describes a two sheeted hyperboloid in $\mathbb{R}^{(1,2)}$ with disconnected parts above and below the $U=0$ plane. The tips of the sheets are located at $(-1,0,0)$ and $(1,0,0)$ in the $(U,X,Y)$ embedding coordinates. The Poincar\'e disk can be obtained by projecting the $U>1$ sheet onto the $U=0$ plane through the point $(-1,0,0)$. In the disk, boundary anchored geodesics are described by the particularly simple equation
\begin{equation}
\tanh{\rho} \ \cos{(\phi-\te)}=\cos \al.
\end{equation}
Here $\te$ denotes the angular coordinate of the center of the geodesic, and $\al\in[0,\pi]$ is the half-opening angle. Pictorially, geodesics in the Poincar\'e disk are arcs of circles that meet the boundary at right angles as in figure \ref{fig:H2coords}. 

Conical defect spacetimes can be obtained as a quotient of AdS$_3$ by identifying surfaces of constant $\phi$ leaving an angular coordinate with a smaller period. In global coordinates the metric is simply
\begin{equation}\label{eqCDmetric}
ds^2=-\cosh^2\rho\ dt^2+d\rho^2+\sinh^2\rho\ d\tilde \phi^2,
\end{equation}
where now $\tilde \phi=\tilde \phi+\frac{2\pi}{N}$. The parameter $N\in(1,\infty)$ gives the strength of the defect. This metric is no longer a solution of the vacuum Einstein equations everywhere but requires a pointlike source at the origin. The defect can be viewed as a static particle of mass $M$ where $4G_{N} M=1-{1}/{N}$. The mass must stay below the black hole limit $M={1}/{4G_{N}}$, which corresponds to $N\to\infty$. For the special cases where $N$ is an integer, the spacetime is a cyclic orbifold AdS$_3/\mathbb{Z}_N$. Some example geodesics in the $t=0$ slice of the conical defect are shown in figure \ref{fig:CDPoincareDiskGeodesics2}, and the corresponding geodesics of AdS$_3$ in figure \ref{fig:PoincareDiskGeodesics}.

The kinematic space corresponding to the Poincar\'e disk was investigated in \cite{Czech2015} and found to be a two dimensional de-Sitter geometry. For ease of comparison the dS$_2$ spacetime can be embedded in the same $\mathbb{R}^{(1,2)}$ where it is a one-sheeted hyperboloid given by
\begin{equation}
-U^2+X^2+Y^2=1.
\end{equation}
The embedding equations 
\begin{equation}
U=\sinh{t},~~X=\cosh t\cos \te,~~Y=\cosh t\sin\te,
\end{equation}
lead to the dS$_2$ metric in global coordinates,
\begin{equation}\label{dS}
ds^2=-dt^2+\cosh^2t \ d\te^2.
\end{equation}
Conformal or ``kinematic" coordinates $(\al,\te)$ will be used often in this paper as they naturally fit with the description of kinematic space as the space of geodesics in AdS$_3$. The transformation $\cosh t=1/\sin\al$, where now $\al\in[0,\pi]$, leads to the dS$_2$ metric
\begin{equation}\label{kinematic}
ds^2=\frac{-d\alpha^2+d\theta^2}{\sin^2\alpha}. 
\end{equation}
With these conventions laid out, the remainder of this section briefly recounts the derivation of the kinematic space geometry for pure AdS$_3$, then details two methods of obtaining the kinematic space for conical defects from the bulk.

\subsection{Kinematic space from differential entropy}\label{secdifferentialentropy}

In \cite{Czech2015} a definition of kinematic space for constant time slices of AdS$_3$ in terms of differential entropy was derived from integral geometry. Each interval of the boundary, denoted by an ordered pair of points $(u,v)$, corresponds to a point in kinematic space covered by null coordinates $(u,v)$. The kinematic space metric in these coordinates was found to be
\begin{equation}\label{eqCrofton}
ds^2=\frac{\pa^2 S(u,v)}{\pa u\pa v} \ du dv,
\end{equation}
where $S(u,v)$ was the length of the shortest oriented geodesic connecting the ends of the interval $(u,v)$ through the bulk. Since the length of a minimal geodesic is holographically interpreted as the entanglement entropy of the interval it subtends, the quantity $\pa^2 S/\pa u\pa v$ was dubbed differential entropy \cite{Ryu2006,Balasubramanian2014,Myers2014,Espindola2017}. However, many interesting spacetimes including the conical defects and BTZ black holes have multiple geodesics connecting pairs of spacelike separated boundary points. Non-minimal geodesics do not correspond to entanglement between spatial regions, but have been conjectured to describe correlations between internal degrees of freedom \cite{Balasubramanian2015}. Because of this potential interest, and their importance in the geodesic approximation for correlation functions \cite{Balasubramanian1999,Asplund2015}, in this paper the differential entropy definition will be expanded to include non-minimal geodesics.

For the constant time slice of AdS$_3$, there is a unique oriented geodesic connecting each ordered pair of boundary points so the issue of non-minimal geodesics in eq. \eqref{eqCrofton} does not arise. Geodesics can be labelled by their half-opening angle $\al$ and centre angle $\te$, and have length
\begin{equation}\label{eqAdSgdlength}
S(\al)=\frac{1}{2G_N}\log \frac{2 \sin\al}{\mu},
\end{equation}
where $\mu$ serves as a gravitational infrared cutoff \cite{Czech2014}. By transforming between kinematic coordinates and null coordinates using $u=\te-\al$, and $v=\te+\al$, eq. \eqref{eqCrofton} can be applied to find
\begin{align}\label{eqds2KS}
\begin{aligned}
ds^2&=\frac{1}{8 G_N}\frac{1}{\sin^2[(v-u)/{2}]} \ du dv\\
&=\frac{1}{8 G_N}\frac{-d\alpha^2+d\theta^2}{\sin^2\alpha}.
\end{aligned}
\end{align}
\begin{figure}
	\centering
        \includegraphics[width=0.48\textwidth]{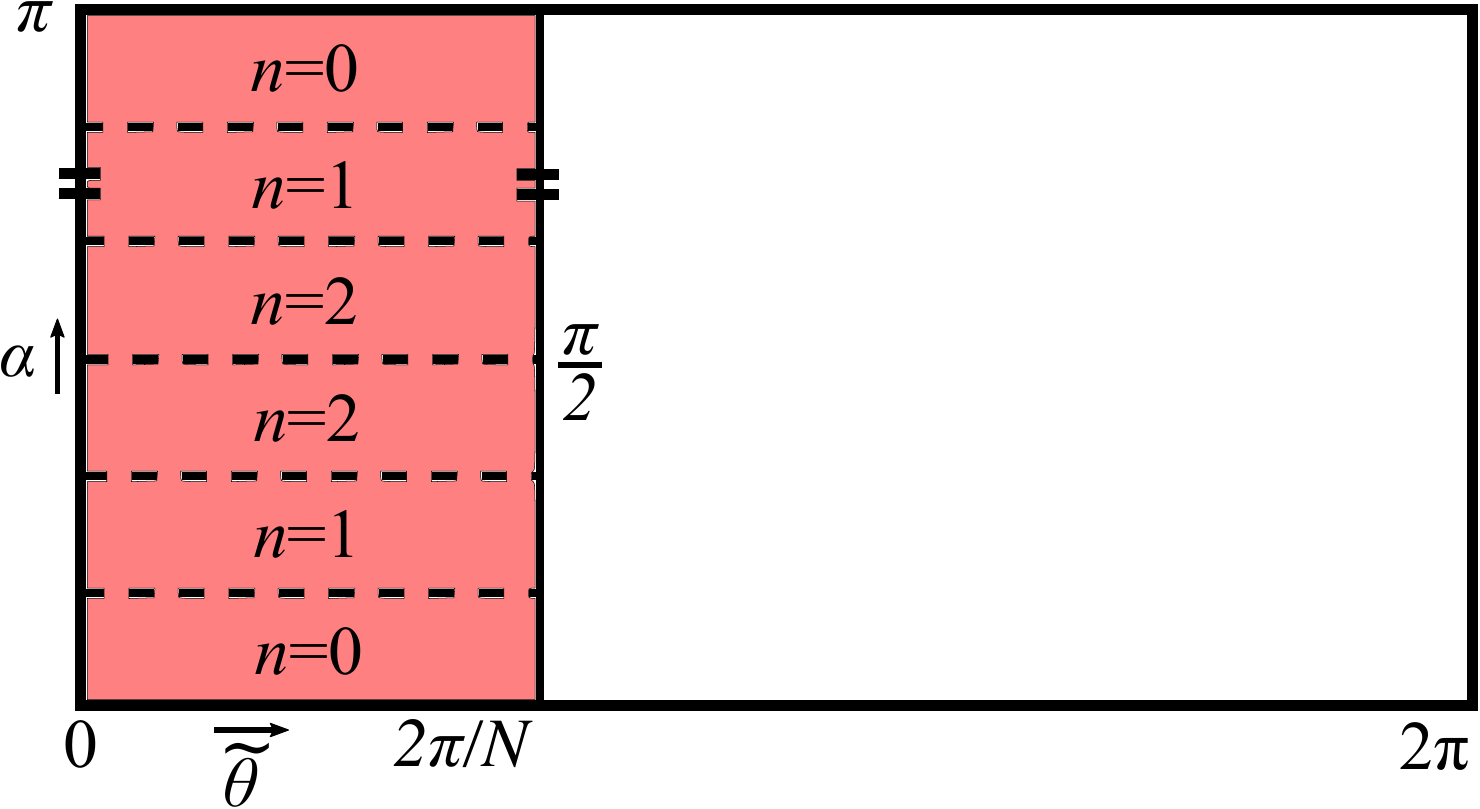}
    \caption{The Penrose diagram for the dS$_2$ kinematic space for pure AdS$_3$ is shown as the full rectangular region, with $\te=\te+2\pi$. For the conical defect case the differential entropy definition of kinematic space produces a vertical strip subregion. The same angular identification which gives the conical defect from AdS$_3$ also gives the kinematic space. $N=3$ is shown for illustration throughout most of this paper.}\label{fig:KSCDVertical}
\end{figure}

Thus, the kinematic space of a constant time slice of AdS$_3$ is dS$_2$ according to the differential entropy definition, as shown in figure \ref{fig:KSCDVertical}. The $\al<\pi/2$ and $\al>\pi/2$ halves are mapped into one another under orientation reversal which acts as $\al\to \pi-\al$ and $\te\to\te+\pi$. The geodesics with $\al=\pi/2$ cut straight across the Poincar\'e disk and have maximal length.

Now consider a constant time slice of the conical defect geometry eq. \eqref{eqCDmetric}. Since the total angle around the boundary is $2\pi/N$, the centre angle of a geodesic will now be denoted $\tilde \te\in[0,2\pi/N]$. Once again, for any pair of boundary points there is a unique minimal geodesic connecting them through the bulk. Minimal geodesics have half-opening angles in the domain $\al\in[0,\pi/2N]$, and by reversing orientations with $\al\to \pi-\al$ and $\tilde \te\to\tilde \te+\pi/N$, also the domain $\al\in[(2N{-}1)\pi/2N,\pi]$. Minimal geodesics cover the top and bottom regions of kinematic space in figure \ref{fig:KSCDVertical}.

In contrast to AdS$_3$, there can be non-minimal geodesics connecting pairs of boundary points. It will be useful to label geodesics and their corresponding regions in kinematic space by the number of times they wind around the defect, $n$. The cases of integer and non-integer $N$ will be treated separately for clarity.

For integer $N$ there are $N-1$ non-minimal geodesics connecting each pair of boundary points, with winding numbers $1\leq n\leq N-1$. Geodesics with winding number $n$ fill in the regions of kinematic space 
\begin{equation}\label{eqalpharanges}
\al\in \left(\frac{n\pi}{2N},\frac{(n+1)\pi}{2N}\right],\quad  \al\in\left[\frac{(2N-n-1)\pi}{2N}, \frac{(2N-n)\pi}{2N}\right),
\end{equation}
where these domains are related by orientation reversal. The upper and lower halves of kinematic space are divided by geodesics with $\al=\pi/2$ which touch the conical defect. On the covering AdS$_3$ space, these are the straight lines through the origin of the Poincar\'e disk. In total there are $2N$ equally sized regions on kinematic space in the $(\al,\tilde \te)$ coordinates.

For non-integer $N$, the maximally winding geodesics have $n=\left\lfloor N\right\rfloor$ and live near the centre line $\al=\pi/2$. There are fewer maximally winding geodesics than other classes, filling out a truncated region 
\begin{equation}
\al\in\left(\frac{(\left\lfloor N\right\rfloor-1)\pi}{2N},\frac{(\left\lfloor N\right\rfloor+1)\pi}{2N}\right).
\end{equation}
Other winding numbers follow eq. \eqref{eqalpharanges}. Each pair of boundary points is connected by $\left\lfloor N\right\rfloor$ or $\left\lfloor N\right\rfloor-1$ geodesics, depending on their angular separation.

The differential entropy definition eq. \eqref{eqCrofton} can be applied to show that the geometry on kinematic space remains locally dS$_2$ for any $N$. The key fact is that minimal and non-minimal geodesics still have lengths given by eq. \eqref{eqAdSgdlength} \cite{Czech2014}. Treating the types on equal footings from the point of view of kinematic space and using $u=\tilde \te-\al$, $v=\tilde \te+\al$ once again gives\footnote{A previous paper \cite{Asplund2016} describing the kinematic spaces for several locally AdS$_3$ geometries, including conical defects, chose to consider only minimal geodesics, and hence found different kinematic space geometries.}
\begin{equation}\label{eqdiffentKS}
ds^2=\frac{1}{8 G_N}\frac{-d\alpha^2+d{\tilde\theta}^2}{\sin^2\alpha}.
\end{equation}
The kinematic space for a constant time slice of a conical defect has the same dS$_2$ metric as the AdS$_3$ case, but with the angular coordinate identified as $\tilde \te\sim\tilde \te+{2\pi}/{N}$. This was expected since the static conical defects are locally AdS$_3$, only differing by the global identification along the angular coordinate. The identification does not affect the lengths of the remaining geodesics. From the differential entropy perspective, the conical defect kinematic space is found by taking an angular quotient of the AdS$_3$ kinematic space; the same quotient that produces the conical defect from pure AdS$_3$ itself. In the next section we show how the quotient acts on geodesics in the covering space, displaying the inherent ambiguities involved in defining kinematic space.
%

\subsection{Kinematic space from boundary anchored geodesics}\label{secbulkKS}

The bulk calculation of the kinematic space for conical defects is more enlightening when the defects are viewed from the perspective of the covering space, AdS$_3$. In particular, it provides motivation for treating minimal and non-minimal geodesics on equal footing in the definition of kinematic space, since there is no real distinction between the types when viewed in the cover. All spacelike geodesics of the conical defect descend from the covering space; the quotient that produces the conical defect divides geodesics into equivalence classes.

\begin{figure}
    \centering
    \begin{subfigure}[b]{0.3\textwidth}
        \includegraphics[width=0.9\textwidth]{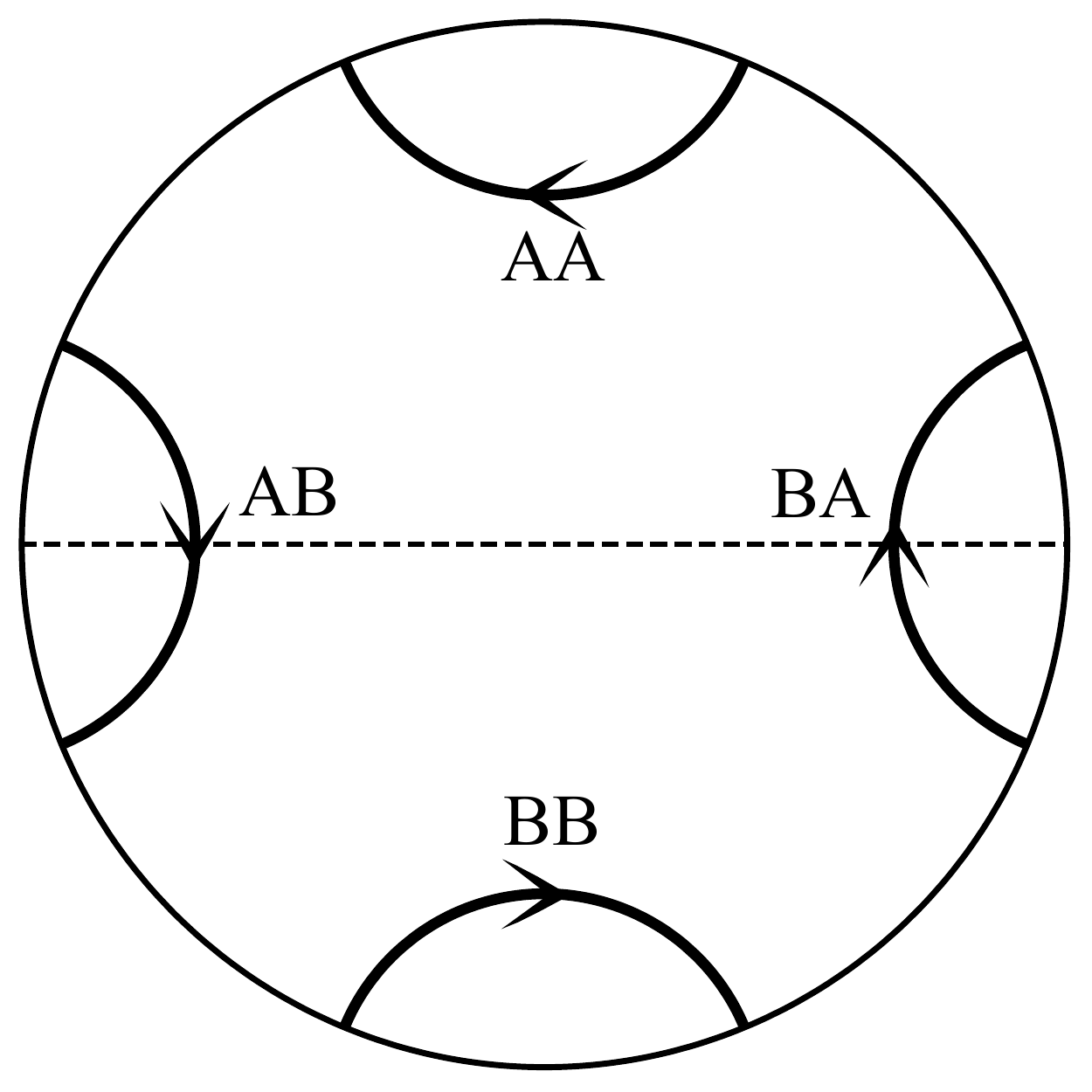}
        \caption{}
        \label{fig:4geodesics}
    \end{subfigure}
     \qquad \qquad
    \begin{subfigure}[b]{0.48\textwidth}
        \includegraphics[width=0.9\textwidth]{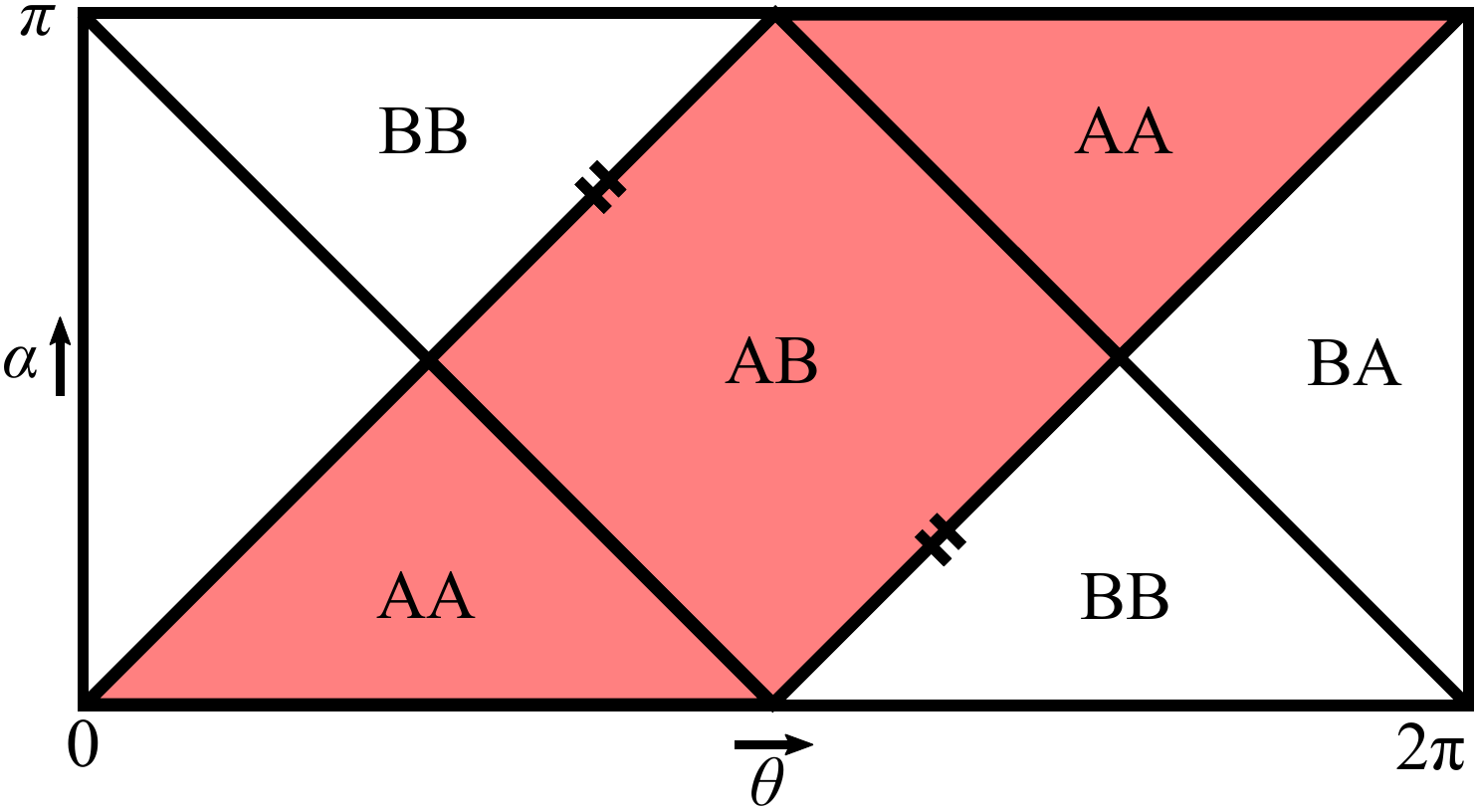}
        \caption{}
        \label{fig:Z2Stripe}
    \end{subfigure}
    \caption{(a) Oriented geodesics in the Poincar\'e disk labelled by their endpoint locations. The $N=2$ wedges are shown with two identified boundaries A and B. (b) Regions of kinematic space labelled by the boundaries each geodesic ends on. For the conical defect with $N=2$, the shaded diagonal strip is a fundamental domain equivalent to the vertical strip.}\label{fig:4}
\end{figure}

As an explicit example, consider the case of $N=2$. The covering space of $\mathbb{H}_2/\mathbb{Z}_2$ is shown in figure \ref{fig:4geodesics}. The covering space can be split into two regions with boundaries labelled $A$ and $B$, which are identified under the quotient. Boundary anchored geodesics on this slice can be grouped into four classes $\{AA,BB,AB,BA\}$ depending on the boundary region their endpoints lie on. The locations of the classes on kinematic space are shown in figure \ref{fig:Z2Stripe}.

Under the $\mathbb{Z}_2$ quotient, $BB$ geodesics are mapped into $AA$ geodesics. Similarly, $BA$ geodesics are mapped into $AB$ geodesics. Therefore, all geodesics in $\mathbb{H}_2/\mathbb{Z}_2$ can be generated by the classes $\{AA,AB\}$, and the $\mathbb{Z}_2$ action. The number of unique geodesics in the conical defect slice is greatly reduced, and similarly for points on kinematic space. As is shown in figure \ref{fig:Z2Stripe}, the kinematic space for the $N=2$ conical defect slice is a diagonal strip of width $\te=\pi$, with the identification $\te= \te+\pi$. However, there are many equivalent ways to choose the fundamental region under the quotient action. If, for example, the classes $\{AA,BA\}$ had been chosen as fundamental, the diagonal strip would point in the opposite direction. Similarly, the entire strip can be shifted by any amount in the $\te$ direction. There is nothing to distinguish these choices, so as in \cite{Zhang2017} a conventional choice has been made.
 
  The quotient only changes the global identification of points in the spacetime, and the geodesics within it. The relationship between nearby geodesics in the kinematic space metric are locally unchanged. While the origin of AdS$_3$ is a fixed point of the quotient, there are no oriented geodesics which are left invariant. From the perspective of kinematic space, the quotient is freely acting, so the metric is expected to be locally unchanged, and the topology to be invariant. This is in contrast to the kinematic space of the BTZ black hole found in \cite{Zhang2017}. The quotient of AdS$_3$ which produces a BTZ black hole has no fixed points so there are no curvature singularities in the BTZ spacetime, but there are geodesics which are fixed under the quotient which changes the topology of kinematic space from a single cylinder to two.
 
  For the more general case of a $\mathbb{Z}_N$ quotient, there are $N^2$ distinct classes of oriented geodesics from the number of ways we can choose two ordered endpoints. The number of distinct regions in kinematic space is $N(N+1)$, one for each of the $N^2$ classes, and one extra for each of the $N$ boundaries. The fundamental region is a diagonal strip with width given by $2\pi/N$ since $\te=\te+2\pi/N$ is identified. This also describes the fundamental region for arbitrary $N$.

 The two approaches presented here, using the differential entropy definition eq. \eqref{eqCrofton}, and studying how the quotient identifies geodesics on the covering space both produce a locally dS$_2$ spacetime but naturally pick out different regions of the kinematic space. The differential entropy definition picks out a vertical strip, while the classification of endpoints on the covering space produces a diagonal strip. However, it is clear from the latter approach that there are many equivalent choices of fundamental region, each with its own merits. The diagonal choice contains some geodesics which have boundary position $\te>2\pi/N$ on the cover. The vertical choice only contains geodesics which are centred at boundary coordinates $\te<2\pi/N$. Since it is easiest to label geodesics with kinematic coordinates $\al\in(0,\pi)$ and $\tilde\te\in[0,2\pi/N]$, the vertical strip will be used in the rest of this paper. 
 
\section{Kinematic space from the boundary}\label{secboundary}

\subsection{Kinematic space metric from conformal symmetry}

In \cite{Czech2016} a definition of kinematic space from the boundary theory was given: each point in kinematic space corresponds to an ordered pair of CFT points.\footnote{In \cite{Czech2016} and \cite{Boer2016} it was shown that an equivalent definition can be made in terms of boundary causal diamonds.} For pure AdS$_3$/CFT$_2$ restricted to a time slice, each ordered pair of CFT points singles out a unique spacelike boundary anchored geodesic so this definition is entirely natural. In the full time dependent geometry, conformal symmetry alone fixes the metric on kinematic space to be
\begin{equation}\label{eqdSdSmetric}
ds^2=4\frac{I_{\mu\nu}(x_1-x_2)}{|x_1-x_2|^2} \ dx_1^\mu dx_2^\nu,
\end{equation}
where 
\begin{equation}
I_{\mu\nu}(x_1-x_2)=\eta_{\mu\nu}-2\frac{(x_1-x_2)_\mu(x_1-x_2)_\nu}{(x_1-x_2)^2},
\end{equation}
is the inversion tensor. The numerical prefactor in the metric is chosen by convention. The two CFT points $x_1^\mu$ and $x_2^\mu$ form a pair of lightlike coordinates on kinematic space with the strange signature $(2,2)$. Since kinematic space is not to be viewed as a physical space, but only as a useful auxiliary space for translating between the bulk and boundary, this is not a concern.

In order to get back the dS$_2$ metric found from the bulk, it is easiest to perform a coordinate transformation from the planar set $x_1^\mu=\{t_1,x_1\}$ to kinematic coordinates on the cylinder. The two pairs of kinematic coordinates are defined through
\begin{align}\label{eq4kinecoords}
\begin{aligned}
\tan \al&=\frac{1}{2}\left(t_1-t_2+x_1-x_2\right),\quad \te=\frac{1}{2}\left(t_1+t_2+(x_1+x_2)\right),\\
\tan \bar\al&=\frac{1}{2}\left(t_1-t_2-(x_1-x_2)\right),\quad \bar\te=\frac{1}{2}\left(t_1+t_2-(x_1+x_2)\right).
\end{aligned}
\end{align}
In terms of these coordinates the kinematic space metric is two copies of the dS$_2$ metric in eq. \eqref{kinematic},
\begin{equation}\label{eqds2ds2}
ds^2=\frac{-d\alpha^2+d\theta^2}{2\sin^2\alpha}+\frac{-d\bar\alpha^2+d\bar \theta^2}{2\sin^2\bar\alpha}.
\end{equation}
Thus the kinematic space for global AdS$_3$, and the dual vacuum state of a CFT$_2$ is dS$_2\times$dS$_2$. When we restrict to a constant time slice by setting $t_1=t_2=0$ we see from eq. \eqref{eq4kinecoords} that $\bar \al$ and $\bar \te$ become redundant coordinates fixed in terms of $\{\al,\te\}$, and that eq. \eqref{eqds2ds2} becomes eq. \eqref{eqdiffentKS}, up to the arbitrarily chosen prefactor.

When the bulk spacetime has non-minimal geodesics, there is no longer a one-to-one correspondence between pairs of CFT points and bulk geodesics. In such a case the argument above cannot be applied. In order to reproduce the quotient structure of kinematic space for conical defects seen in section \ref{secbulk}, another approach must be taken. We take the point of view espoused in \cite{Karch2017}; OPE blocks in the CFT should be viewed as free fields on kinematic space, and their equation of motion reflects the geometry of kinematic space.

\subsection{OPE blocks}\label{secOPEblocks}

In \cite{Czech2016}, the operator product expansion (OPE) of two scalar CFT operators was broken into OPE blocks, and these blocks were identified as fields on kinematic space. Two scalar operators $\op_i(x_1)$ and $\op_j(0)$ in a planar CFT with conformal weights $\De_i$ and $\De_j$ respectively can be expanded in terms of a local basis of operators at the origin,
\begin{equation}\
\op_{i}(x)\op_{j}(0) =  \sum_{k}C_{ijk}\left|x\right|^{\Delta_{k}-\Delta_{i}-\Delta_{j}}\big(1+b_{1}\,x^{\mu}\partial_{\mu}+b_{2}\,x^{\mu}x^{\nu}\partial_{\mu}\partial_{\nu}+\ldots\big)\op_{k}(0).
\end{equation}
This is the OPE, where the quasiprimaries $\op_k(0)$, and their descendants given by the derivative terms, form the basis of operators at the origin. Notably, the $b_n$ coefficients are completely fixed by conformal symmetry, while the $C_{ijk}$ are simply constants, but are theory-dependent. Each term in the sum has a characteristic scaling dimension $\De_k$, the dimension of the quasiprimary $\op_k$, and represents the contribution to the OPE of the entire conformal family of $\op_k$. Each of these terms can be packaged into a new operator $\mathcal{B}^{ij}_k(x_1,x_2)$ called an OPE block, and the OPE can be written as
\begin{equation}\label{eqOPEblock}
\op_{i}\left(x_1\right)\op_{j}\left(x_2\right) = \left|x_1-x_2\right|^{-\Delta_{i}-\Delta_{j}}\sum_{k}C_{ijk}\mathcal{B}_{k}^{ij}\left(x_1,x_2\right).
\end{equation}

Since the OPE blocks depend on a pair of CFT points, the two points where operators in the OPE are inserted, it is natural to view them as fields on kinematic space. A major insight of \cite{Czech2016} was that the Casimir eigenvalue equation satisfied in the CFT by the OPE blocks can be interpreted as a wave equation. The differential representation of the CFT Casimir operator appropriate for OPE blocks is the Laplacian in the kinematic space metric eq. \eqref{eqds2ds2}. This gives yet another prescription for determining the kinematic space for a CFT state which is applicable when arguments from conformal symmetry alone are not sufficient, as advocated for recently in \cite{Karch2017}. In the following section, we will show how this prescription can be modified and used to obtain the kinematic space for excited CFT states dual to conical defects, in agreement with the results of section \ref{secbulk}. First, we review how the bulk metric of AdS$_3$ can be determined from a quadratic CFT$_2$ Casimir in a differential representation appropriate for scalar fields, and how the bilocal scalar representation of OPE blocks gives the metric on kinematic space. These initial cases have been summarized in \cite{Czech2016,Boer2016}.

In a 2d CFT, the global conformal group $SO(2,2)$ forms a subgroup of the larger Virasoro symmetry group. The global subgroup corresponds holographically to the isometries of pure AdS$_3$ with appropriate boundary conditions, while the other generators of the Virasoro group are associated to transformations which preserve the asymptotic boundary \cite{Brown1986}. The global conformal generators $L_{0,\pm1}, \ \bar L_{0,\pm 1}$ in the standard basis satisfy two copies of the Witt algebra
\begin{equation}\label{Witt}
[L_n,L_m]=(n-m)L_{n+m},\quad [\bar L_n,\bar L_m]=(n-m)\bar L_{n+m}, \quad [L_n,\bar L_m]=0.
\end{equation}
When acting on conformal operators, the algebra is represented by some differential operators $\mathcal{L}_{n}$ as
\begin{equation}\label{eqdiffrep}
[L_n,\op_k(x)]=\mathcal{L}_{n} \op_k(x),
\end{equation}
which depend on the $SO(2,2)$ representation of $\op_k$. 

The quadratic Casimir operator
\begin{equation}\label{eqc2}
\mathcal{C}_2=-\frac{1}{2}L^{AB}L_{AB}=-2L_0^2+(L_1 L_{-1}+L_{-1}L_1)\ + \ (L\to\bar L),
\end{equation}
commutes with all the global conformal generators.\footnote{Our conventions are as in \cite{Boer2016}.} Here, $L_{AB}$ is written as an $SO(2,2)$ Lorentz operator in the embedding space formalism \cite{Dolan2004}. Quasiprimary operators $\op_k(x)$ are eigenoperators of this Casimir obeying
\begin{equation}\label{eqcas}
[\mathcal{C}_2,\op_k(x)]=-\frac{1}{2}\mathcal{L}^{AB}\mathcal{L}_{AB}\op_k(x)=C_{k} \op_k(x),
\end{equation}
where for a quasiprimary with scaling dimension $\De_k$ and spin $l_k$ the eigenvalue is
\begin{equation}\label{eqcaseival}
C_k=\De_k(\De_k - d)-l_k(l_k+d-2).
\end{equation}
The same eigenvalue applies to the conformal Casimir in higher dimensional CFT's although we only consider $d=2$ here. Since descendants of $\op_k(x)$ are obtained through the action of conformal generators which commute with $\mathcal{C}_2$, descendants obey the same Casimir eigenvalue equation. Thus, Casimir eigenvalues classify irreducible representations of the global conformal group.

The holographic interpretation of the Casimir equation \eqref{eqcas} depends on the representation used for the conformal generators. As an example, consider a scalar quasiprimary operator $\op_k$ with dimension $\De_k$, dual to a massive bulk scalar field $\varphi$. In terms of right and left moving planar CFT coordinates $\xi=x+t$, $\bar \xi=x-t$, the appropriate differential representation of the global conformal generators is
\begin{equation}
\mathcal{L}_{-1}=\pa_\xi,\quad \mathcal{L}_0=-\xi\pa_\xi-\frac{1}{2}\De_k,\quad \mathcal{L}_1=\xi^2\pa_\xi+\xi\De_k,
\end{equation}
and similarly for barred generators with $\xi\to\bar\xi$. An explicit calculation of eq. \eqref{eqcas} using eq. \eqref{eqc2} verifies that $[\mathcal{C}_2,\op_k(x)]=\De_k(\De_k - 2)\op_k(x)$. 

Holographically, the global conformal generators correspond with AdS$_3$ isometries. Scale/radius duality prescribes that the scaling dimension $\De_k$ be replaced by the radial scale operator $z\pa_z$. Then the conformal generators become
\begin{equation}
\eta_{-1}=\pa_\xi,\quad \eta_0=-\xi\pa_\xi-\frac{1}{2}z\pa_z,\quad \eta_1=\xi^2\pa_\xi+\xi z\pa_z,
\end{equation}
with a barred sector given by $\xi\to\bar\xi$. These operators still satisfy the algebra eq. \eqref{Witt} under the Lie bracket. However, this algebra now admits a non-trivial extension 
\begin{equation}
\eta_1\to\xi^2\pa_\xi+\xi z\pa_z-z^2\pa_{\bar\xi},\quad \bar\eta_1\to\bar\xi^2\pa_{\bar\xi}+\bar\xi z\pa_z-z^2\pa_{\xi},
\end{equation}
which leaves the Lie brackets between all elements unchanged, and which vanishes in the boundary limit $z\to0$. Using the extended algebra, and replacing $\op_k$ by its dual field, the Casimir equation \eqref{eqcas} becomes
\begin{equation}\label{eqcasadslaplacian}
\left(z\pa_z-z^2\pa_z^2-4z^2\pa_\xi \pa_{\bar\xi}\right)\varphi=-\square_{\mathrm{AdS}}\varphi=-m^2\varphi,
\end{equation}
which is the Klein-Gordon equation for a massive scalar field in Poincar\'e AdS$_3$, with $m^2=-\De_k(\De_k-2)$ \cite{Maldacena1998}. In the $\De_k$ scalar representation, the global conformal Casimir can be identified as the AdS$_3$ Laplacian, $\mathcal{C}_2=-\square_{\mathrm{AdS}}$. 

In a similar manner, the Laplacian for the kinematic space of the CFT$_2$ vacuum state can be derived from the Casimir in an appropriate representation. The authors of \cite{Czech2016} identified this representation from the transformation properties of OPE blocks, the  natural candidates for fields on kinematic space. Under a conformal transformation, a spin-zero local operator with scaling dimension $\De_i$ transforms as
\begin{equation}
\op_i\left(x\right)\to \Om\left(x'\right)^{\De_i} \op_i\left(x'\right),\quad \Om\left(x'\right)=\det\left(\frac{\pa x'^\mu}{\pa x^\nu}\right),
\end{equation}
while
\begin{equation}
\left|x_1-x_2\right|\to\left(\Om\left(x_1'\right)\Om\left(x_2'\right)\right)^{-1/2}\left|x_1'-x_2'\right|.
\end{equation}
From eq. \eqref{eqOPEblock}, these transformation laws imply that OPE blocks obey
\begin{equation}
\mathcal{B}_{k}^{ij}\left(x_1,x_2\right)\to\left(\frac{\Om(x_1')}{\Om(x_2')} \right)^{(\De_i-\De_j)/2}\mathcal{B}_{k}^{ij}\left(x_1',x_2'\right).
\end{equation}
Restricting to the case of $\De_i=\De_j$ shows that the equal-weight OPE block transforms in a spinless, $\De=0$ representation in each of its coordinates. This is the same transformation law as a pair of dimensionless scalar operators $\varphi_1(x_1)\varphi_2(x_2)$. The action of the conformal generators on this pair is, from eq. \eqref{eqdiffrep},
\begin{align}
\begin{aligned}
{[L_n,\varphi_1(x_1)\varphi_2(x_2)]}&=[L_n,\varphi_1(x_1)]\varphi_2(x_2)+\varphi_1(x_1)[L_n,\phi_2(x_2)]\\
&=(\mathcal{L}_{n,1}+\mathcal{L}_{n,2})\varphi_1(x_1)\varphi_2(x_2),
\end{aligned}
\end{align}
where $\mathcal{L}_{n,k}$ is the $\De=0,l=0$ differential representation of $L_n$ acting only on the $x_k$ coordinates. The OPE block is a linear combination of a single quasiprimary and its descendants, so it satisfies a Casimir eigenvalue equation with the same eigenvalue \eqref{eqcaseival} as the quasiprimary,
\begin{equation}\label{eqbilocalcas}
[\mathcal{C}_2,\mathcal{B}_{k}\left(x_1,x_2\right)]=-\frac{1}{2}(\mathcal{L}^{AB}_1+\mathcal{L}^{AB}_2)(\mathcal{L}_{AB,1}+\mathcal{L}_{AB,2})\mathcal{B}_{k}\left(x_1,x_2\right)=C_k\mathcal{B}_{k}\left(x_1,x_2\right).
\end{equation}
Employing an explicit representation for the conformal generators will produce a differential equation for the OPE blocks which can be interpreted as a Klein-Gordon equation on kinematic space.

From the global AdS$_3$ Killing vectors
\begin{equation}\label{eqKV}
\small
\begin{split}
\xi_{-1}&=\frac{1}{2}e^{-i(t+\phi)}(\tanh(\rho)\pa_t +i\pa_\rho+\coth(\rho)\pa_\phi),\\
\xi_0&=\frac{1}{2}(\pa_t+\pa_\phi),\\
 \xi_1&=\frac{1}{2}e^{i(t+\phi)}(\tanh(\rho)\pa_t -i\pa_\rho+\coth(\rho)\pa_\phi),\\
\end{split}
\quad \quad 
\begin{split}
\bar\xi_{-1}&=\frac{1}{2}e^{-i(t-\phi)}(\tanh(\rho)\pa_t +i\pa_\rho-\coth(\rho)\pa_\phi),\\
\bar\xi_0&=\frac{1}{2}(\pa_t-\pa_\phi),\\
 \bar \xi_1&=\frac{1}{2}e^{i(t-\phi)}(\tanh(\rho)\pa_t -i\pa_\rho-\coth(\rho)\pa_\phi),
\end{split}
 \normalsize
\end{equation}
we can obtain a differential representation of the conformal generators on the cylinder by taking the $\rho\to\infty$ boundary limit \cite{Brown1986}, 
\begin{equation}\label{eqcylgen}
\begin{split}
    \mathcal{L}_{-1}&=\frac{1}{2}e^{-i(t+\phi)}(\pa_t +\pa_\phi),\\
\mathcal{L}_0&=\frac{1}{2}(\pa_t+\pa_\phi),\\
 \mathcal{L}_1&=\frac{1}{2}e^{i(t+\phi)}(\pa_t +\pa_\phi),\\
  \end{split}
  \quad \quad 
  \begin{split}
    \bar{\mathcal{L}}_{-1}&=\frac{1}{2}e^{-i(t-\phi)}(\pa_t -\pa_\phi),\\
\bar{\mathcal{L}}_0&=\frac{1}{2}(\pa_t-\pa_\phi),\\
 \bar{\mathcal{L}}_1&=\frac{1}{2}e^{i(t-\phi)}(\pa_t -\pa_\phi).
  \end{split}
\end{equation}
Using this representation to calculate the Casimir in its bilocal scalar representation \eqref{eqbilocalcas} requires computing
\begin{equation}\label{eqbilocalcas2}
-\frac{1}{2}\mathcal{L}_{AB,1}\mathcal{L}^{AB}_1-\frac{1}{2}\mathcal{L}_{AB,2}\mathcal{L}^{AB}_2+\mathcal{L}_{AB,1}\mathcal{L}^{AB}_2,
\end{equation}
as in eq. \eqref{eqc2}. This task is simplified since the two terms which act on only a single coordinate do not contribute. This can be verified directly from the representation \eqref{eqcylgen}, or by noting that $L_{AB,i}L^{AB}_i$ acting on $\mathcal{B}_{k}\left(x_1,x_2\right)$ produces the eigenvalue \eqref{eqcaseival}, which vanishes for the $\De=0,\ l=0$ representation appropriate for the equal-weight OPE blocks in $d=2$. 

The term with mixed derivatives does not vanish. It is
\begin{equation}
\mathcal{L}_{AB,1}\mathcal{L}^{AB}_2=-4\left(\bar{\mathcal{L}}_{0,1}\bar{\mathcal{L}}_{0,2}+\mathcal{L}_{0,1} \mathcal{L}_{0,2}\right)+2\left[\bar{\mathcal{L}}_{-1,1}\bar {\mathcal{L}}_{1,2}+\mathcal{L}_{1,1}\mathcal{L}_{-1,2}+\bar{\mathcal{L}}_{1,1}\bar{\mathcal{L}}_{-1,2}+\mathcal{L}_{-1,1}\mathcal{L}_{1,2}\right],
\end{equation}
where the second index indicates which point in the pair $(x_1,x_2)$ the operator acts on. Using eq. \eqref{eqcylgen} leads to 
\begin{align}
\begin{aligned}
\mathcal{L}_{AB,1}\mathcal{L}^{AB}_2=-2\left(\pa_{t_1}\pa_{t_2}+\pa_{\phi_1}\pa_{\phi_2}\right)&+\cos\left(t_1-t_2+\phi_1-\phi_2\right)\left(\pa_{t_1} +\pa_{\phi_1}\right)\left(\pa_{t_2} +\pa_{\phi_2}\right)\\
&+\cos\left(t_1-t_2-\left(\phi_1-\phi_2\right)\right)\left(\pa_{t_1} -\pa_{\phi_1}\right)\left(\pa_{t_2} -\pa_{\phi_2}\right).
\end{aligned}\end{align}
This operator simplifies greatly if we introduce coordinates analogous to the kinematic coordinates used in eq. \eqref{eq4kinecoords},\footnote{There is no longer a $\tan$ because this transformation is between sets of coordinates on the cylinder.}
\begin{equation}\label{eqcombinedtrans}
\begin{split}
    \al&=\frac{1}{2}\left(t_1-t_2+(\phi_1-\phi_2)\right),\\
\bar \al&=\frac{1}{2}\left(t_1-t_2-(\phi_1-\phi_2)\right),\\
  \end{split}
  \quad \quad 
  \begin{split}
    \te&=\frac{1}{2}\left(t_1+t_2+\phi_1+\phi_2\right),\\
\bar \te&=\frac{1}{2}\left(t_1+t_2-(\phi_1+\phi_2)\right),\\
  \end{split}
\end{equation}
which leads to
\begin{align}
\begin{aligned}
\mathcal{L}_{AB,1}\mathcal{L}^{AB}_2=-2\sin^2 \al\left(-\pa_\al^2+\pa_\te^2\right)-2\sin^2\bar \al\left(-\pa_{\bar \al}^2+\pa_{\bar \te}^2\right).
\end{aligned}
\end{align}
The Casimir equation for the OPE block is then
\begin{align}\label{eqopecas}
\begin{aligned}
\small
{\left[\mathcal{C}_2,\mathcal{B}_{k}\left(x_1,x_2\right)\right]}&=\left[-2\sin^2 \al \left(-{\pa_\al^2}{+}\pa_\te^2\right)-2\sin^2\bar \al\left(-{\pa_{\bar \al}^2}{+}\pa_{\bar \te}^2\right)\right]\mathcal{B}_{k}\left(x_1,x_2\right)=\De_k({\De_k}{-}2)\mathcal{B}_{k}.
\normalsize
\end{aligned}
\end{align}
It is easy to check that this operator is the scalar Laplacian in the dS$_2\times$dS$_2$ metric \eqref{eqds2ds2} found from conformal symmetry arguments. This motivates the interpretation of an OPE block as a negative mass scalar field propagating freely on kinematic space \cite{Czech2016},
\begin{equation}\label{eqOPEblockeom}
\left(\square_{dS}+\bar{\square}_{dS}\right)\mathcal{B}_{k}\left(x_1,x_2\right)=m^2\mathcal{B}_{k},
\end{equation}
with the mass term $m^2=-\De_k(\De_k-2)$ given by the Casimir eigenvalue \eqref{eqcaseival} for the quasiprimary of the block. Again, kinematic space is meant to be a useful auxiliary space, not a physical one, so the appearance of negative mass fields is not a concern.

In the following section the equal-time OPE will be considered for CFT's dual to conical defects. Setting $t_1=t_2=0$ in eq. \eqref{eqcombinedtrans} and eliminating two redundant coordinates in eq. \eqref{eqopecas} leads to the Laplacian for a single dS$_2$ spacetime,
\begin{align}\label{eqequaltimelaplacian}
\begin{aligned}
{[\mathcal{C}_2,\mathcal{B}_{k}\left(t=0,\al,\te\right)]}&=-4\sin^2 \al\left(-\frac{\pa^2}{\pa\al^2}+\frac{\pa}{\pa\te^2}\right)\mathcal{B}_{k}\left(t=0,\al,\te\right).
\end{aligned}
\end{align}

\subsection{CFT dual to conical defects}

Conical defect spacetimes can be created by adding a particle to pure AdS and are dual to certain excited states of the boundary theory \cite{Balasubramanian1999,Lunin2003}. The dual CFT is discretely gauged and lives on a cylinder with an angular identification inherited from the bulk. For the conical defects with integer $N$ it is often useful to consider a covering CFT living on the boundary of pure AdS$_3$ that ungauges the discrete $\mathbb{Z}_N$ symmetry \cite{Balasubramanian2015}.\footnote{The covering CFT only inherits a Virasoro symmetry group when $N$ is an integer \cite{Boer2011}.} Physical, gauge invariant quantities in the base CFT can be computed from appropriately symmetrized quantities on the cover. This method of images on the cover is a common way to calculate correlation functions of operators in the base CFT \cite{Balasubramanian1999,Balasubramanian2003,Arefeva2016,Arefeva2016a,Arefeva2017}. It is important to note that the covering CFT is not identical to the base CFT, as there are many non-symmetrized quantities on the cover that do not correspond to physical, gauge invariant quantities on the base. In addition, the two theories do not share the same central charge. In line with section \ref{secbulk}, quantities on the base where $\tilde\phi\in[0,2\pi/N]$ will be marked with a tilde to distinguish them from quantities on the cover where $\phi\in[0,2\pi]$.

Restricting to integer $N$, a base operator $\tilde\op(t,\tilde\phi)$ of dimension $\De$ can be represented on the cover by a symmetrized operator
\begin{equation}\label{eqsym}
\tilde\op\left(t,\tilde\phi\right)=\frac{1}{N}\sum_{m=0}^{N-1} \exp{\left(i\frac{2\pi m}{N}\frac{\pa}{\pa \phi}\right)}\op\left(t,\phi\right),
\end{equation}
where $\op(t,\phi)$ is an operator on the cover of the same dimension $\De$, with $ \phi\in[0,2\pi]$, and the first copy $(m=0)$ is inserted at $\phi=\tilde\phi$ by convention.\footnote{The equality of scaling dimensions here is a consequence of unitarity in a 1+1d CFT and may not be guaranteed in higher dimensions.} This convention is somewhat arbitrary. It reflects the freedom to choose a fundamental domain on the kinematic space, as will become clear. The timelike coordinates of the two theories are simply identified, and we work on a fixed time slice in both cases. The generators of rotation
\begin{equation}
\exp{\left(i\frac{2\pi m}{N}\frac{\pa}{\pa\phi}\right)},
\end{equation}
are conformal generators that have the effect of permuting through copies of $\op(t,\phi)$ equally spaced around the circle. An equivalent expression to eq. \eqref{eqsym} is
\begin{equation}
\tilde\op\left(t,\tilde\phi\right)=\frac{1}{N}\sum_{m=0}^{N-1} \op\left(t,\phi+\frac{2\pi m}{N}\right).
\end{equation}

\subsubsection{Partial OPE block decomposition}

The main goal of this section will be to obtain a symmetrized expression for the equal-time base OPE in terms of cover OPE blocks. That expression can then be used to determine the appropriate Casimir eigenvalue equation for the blocks, and in turn the kinematic space geometry can be inferred. The base OPE of equal-time operators inserted at locations $(t,\tilde\phi_1)$ and $(t,\tilde\phi_2)$ with $\tilde\phi_1>\tilde\phi_2$ is of the form

 \begin{equation}\label{equsualOPE}
\tilde\op_{i}\left(t,\tilde\phi_1\right)\tilde\op_{j}\left(t,\tilde\phi_2\right) = \left|2-2\cos(\tilde \phi_1-\tilde\phi_2)\right|^{-\Delta}\sum_{k}\tilde C_{ijk}\tilde{\mathcal{B}}_{k}\left(t,\tilde\phi_1,\tilde\phi_2\right),
\end{equation}
where $\tilde{\mathcal{B}}_{k}\left(t,\tilde\phi_1,\tilde\phi_2\right)$ are the equal-time base OPE blocks.\footnote{Here, operators on the cylinder have been rescaled relative to the planar operators used in section \ref{secOPEblocks}, see \cite{Pappadopulo2012} for example. In the OPE limit $\tilde\phi_2\to\tilde\phi_1$ where the curvature of the cylinder becomes unimportant, one recovers the form of eq. \eqref{eqOPEblock} for a planar CFT.}   Again, $\De_i=\De_j=\De$ so the indices $i,j$ on the OPE blocks are dropped for brevity. The base OPE can be rewritten using eq. \eqref{eqsym}, after which the OPE between cover operators can be broken into OPE blocks to get
\begin{align}
\begin{aligned}\label{eqdoublesumexpansion}
\tilde{\mathcal{O}}_i(t,  \tilde\phi_1) \tilde{\mathcal{O}}_j(t,  \tilde\phi_2)  =& \frac{1}{N^2}\sum_{a=0}^{N-1} \sum_{b=0}^{N-1}  \exp{\left(i\frac{2\pi a}{N}\frac{\pa}{\pa \phi_1}\right)}  \exp{\left(i\frac{2\pi b}{N}\frac{\pa}{\pa  \phi_2}\right)}  {\mathcal{O}}_i(t, \phi_1) {\mathcal{O}}_j(t, \phi_2) \\
=& \frac{1}{N^2}\sum_{a=0}^{N-1} \sum_{b=0}^{N-1}  \exp{\left(i\frac{2\pi a}{N}\frac{\pa}{\pa \phi_1}\right)}  \exp{\left(i\frac{2\pi b}{N}\frac{\pa}{\pa \phi_2}\right)} \\
&\cdot \left[ \left|2-2\cos(\phi_1-\phi_2)\right|^{-\De}\sum_{k} C_{ijk}{\mathcal{B}}_{k}\left(t, \phi_1,\phi_2\right)\right].
\end{aligned}
\end{align}
The structure constants and OPE blocks may be different on the cover compared to the base, and are differentiated by a tilde. 

 Now in the covering space we introduce kinematic coordinates of the form (cf. \eqref{eqcombinedtrans})
\begin{equation}
\al=\frac{1}{2}(\phi_1-\phi_2),\quad \te=\frac{1}{2}(\phi_1+\phi_2).
\end{equation}
The permutation generators can be rewritten 
\begin{equation}
\exp{\left(i\frac{2\pi a}{N}\frac{\pa}{\pa  \phi_1}\right)}\exp{\left(i\frac{2\pi b}{N}\frac{\pa}{\pa  \phi_2}\right)}=\exp{\left(i\frac{2\pi (a-b)}{N}\frac{\pa}{\pa  \phi_1}\right)}\exp{\left(i\frac{2\pi b}{N}\frac{\pa}{\pa  \te}\right)}.
\end{equation}
The $N^2$ terms in the double sum \eqref{eqdoublesumexpansion} can be reorganized into a more appealing form 
\begin{align}\label{eqphiteexpansion}
\begin{aligned}
\tilde{\mathcal{O}}_i(t, \tilde\phi_1) \tilde{\mathcal{O}}_j(t, \tilde\phi_2&)  =\\ \frac{1}{N^2}\sum_{k} C_{ijk} &\sum_{m=0}^{N-1} \exp{\left(i\frac{2\pi m}{N}\frac{\pa}{\pa \phi_1}\right)}  \left[ \left|2-2\cos( 2\al)\right|^{-\De}\sum_{b=0}^{N-1} \exp{\left(i\frac{2\pi b}{N}\frac{\pa}{\pa  \te}\right)}\mathcal{B}_{k}\left(t,\al,\te\right)\right].
\end{aligned}
\end{align}
The interior sum over $b$ accounts for the $N$ terms where both points $\phi_1$ and $\phi_2$ are shifted by the same amount, that is $a=b$. In this case $\al$ is fixed; the $\pa/\pa{\te}$ generator permutes between images of the pair of operators on the cover. From the bulk viewpoint, $\pa/\pa{\te}$ permutes through the $N$ images of a geodesic that are identified under $\mathbb{Z}_N$. In the exterior sum, the $\pa/\pa{\phi_1}$ generators increase the angular distance between the insertion points. In bulk terms, $\pa/\pa\phi_1$ changes the winding number of geodesics connecting the boundary points. 

\begin{figure}
      \centering
        \includegraphics[width=.3\textwidth]{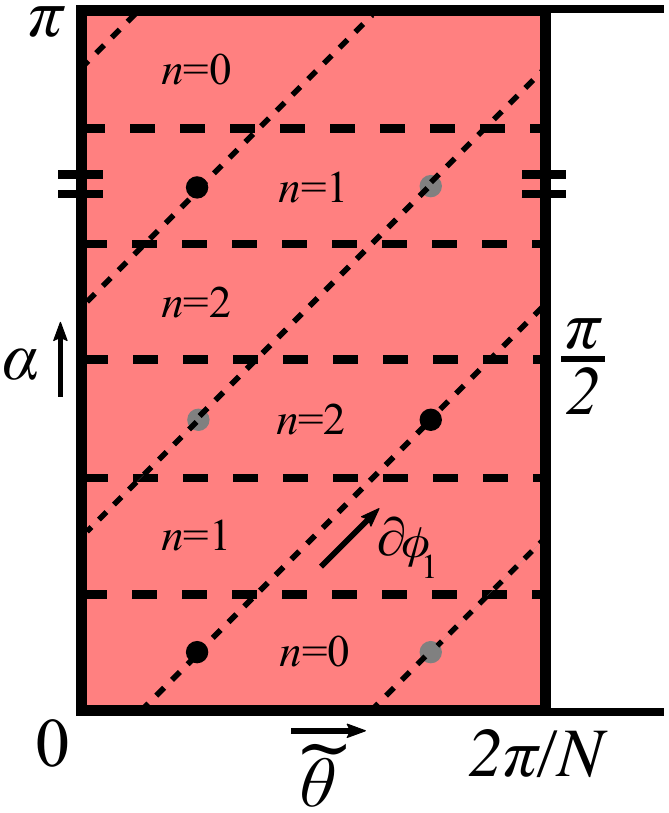}
        \caption{On kinematic space, $\phi_1$ is a null coordinate. In terms of cover operators, acting with $\exp{(i\frac{2\pi}{N}\frac{\pa}{\pa \phi_1})}$ increases the angular separation $\al$. In terms of conical defect geodesics, acting once with $\exp{(i\frac{2\pi}{N}\frac{\pa}{\pa \phi_1})}$ increases the winding number while leaving the endpoints fixed. All winding numbers are reached by acting with $\exp{(i\frac{2\pi}{N}\frac{\pa}{\pa \phi_1})}$ generators, in contrast to $\exp{(i\frac{2\pi}{N}\frac{\pa}{\pa \al})}$ generators. }
        \label{fig:KSCDdiagonalpath}
\end{figure}

It may seem more natural to use $\pa/\pa{ \al}$ generators along with the $\pa/\pa \te$ generators. However, when $N$ is an even integer, acting with $\pa/\pa\al$ alone does not reach images on the cover of all separations $\al$. In bulk terms, not all winding numbers for geodesics with a given orientation can be reached with $\pa/\pa\al$ generators alone. In order to reach all images for all integer $N$, a combination of $\pa/\pa\te$ and one of $\pa/\pa\phi_1$ or $\pa/\pa\phi_2$ is needed, as illustrated in figure \ref{fig:KSCDdiagonalpath}.

  \begin{figure}
    \centering
    \begin{subfigure}[b]{0.3\textwidth}
        \includegraphics[width=\textwidth]{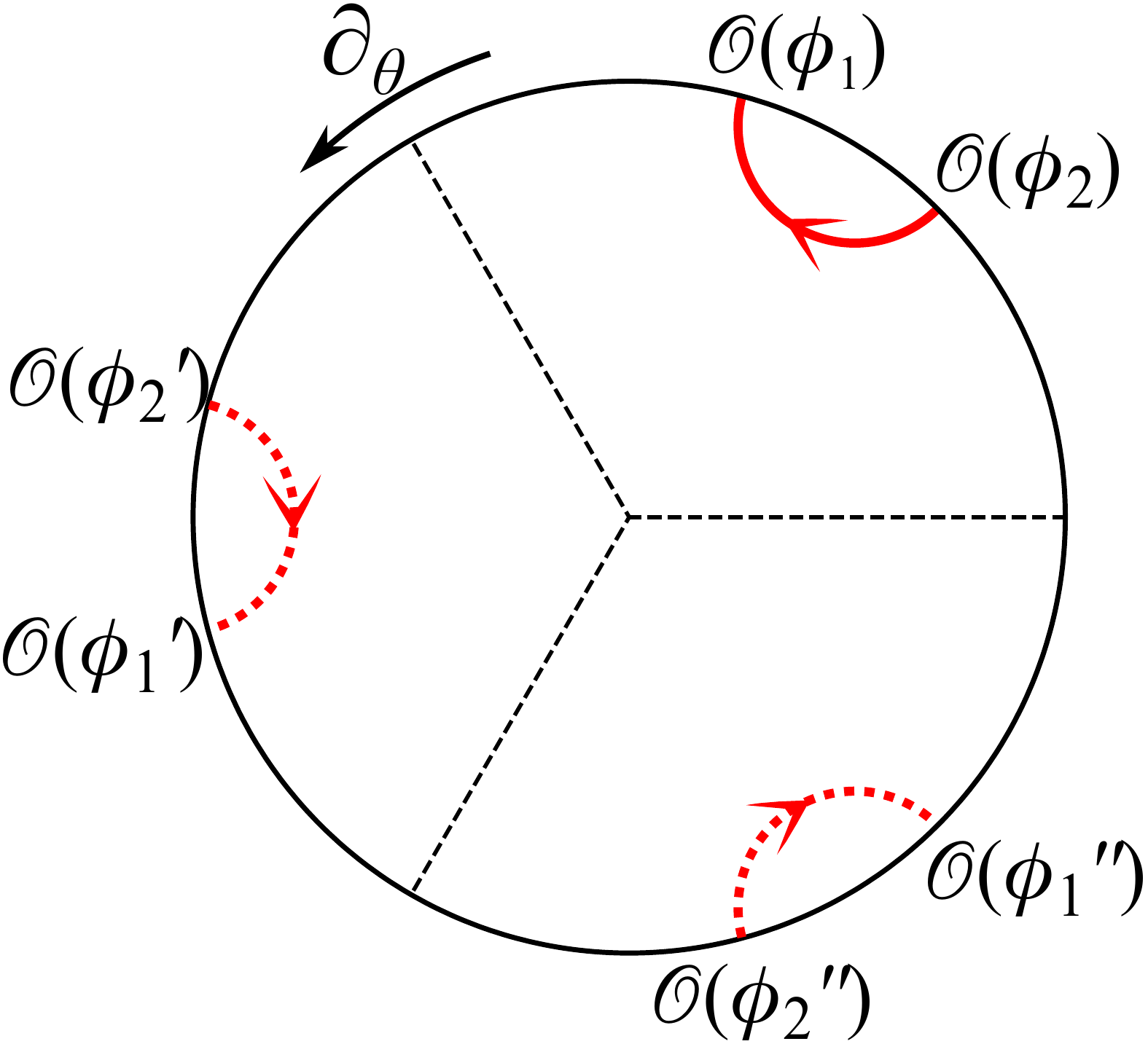}
        \caption{$\mathcal{B}_{k,m=0}$, $n=0$ geodesics.}
        \label{fig:3RedGDs}
    \end{subfigure}
    \quad
    \begin{subfigure}[b]{0.3\textwidth}
        \includegraphics[width=\textwidth]{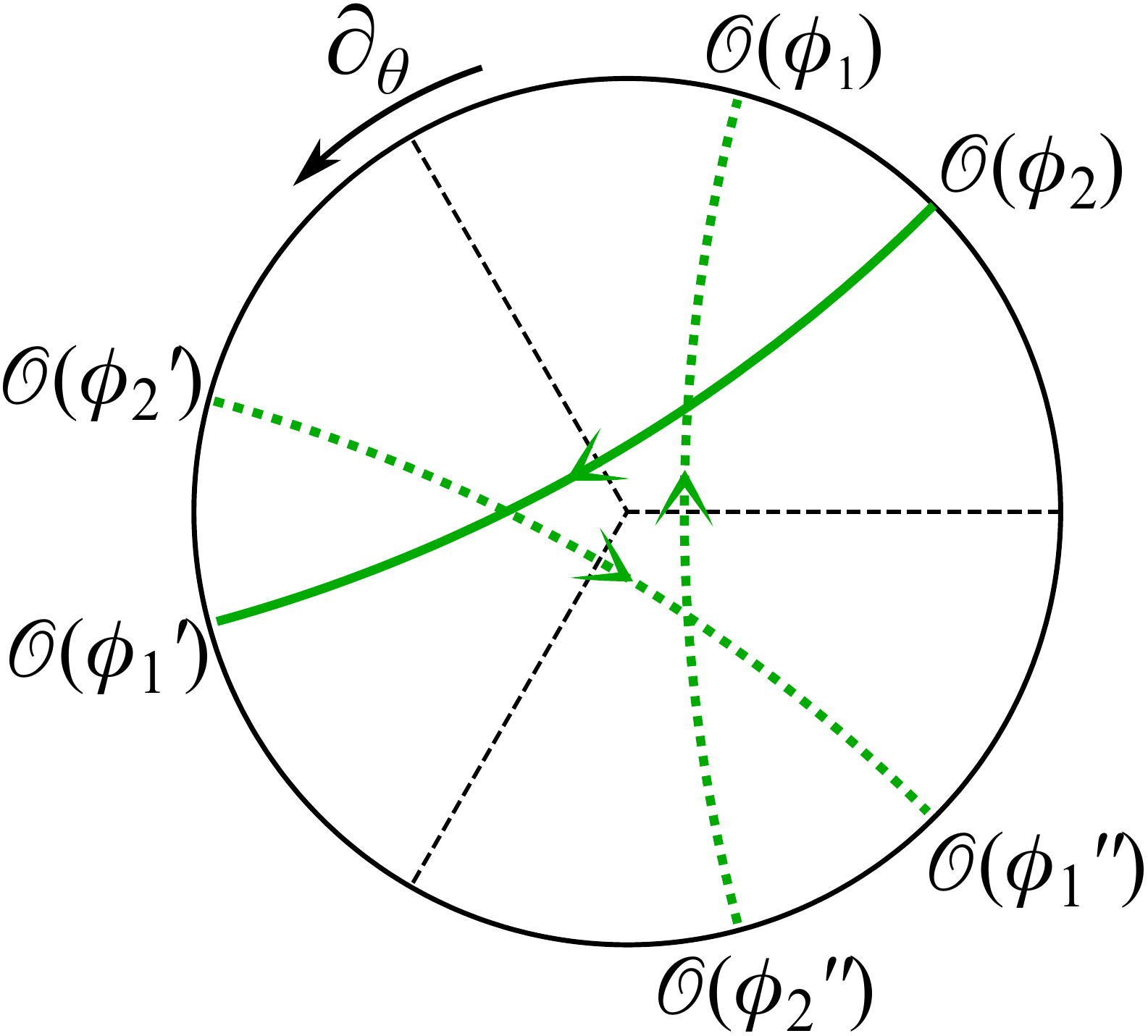}
        \caption{$\mathcal{B}_{k,m=1},~n=2$~geodesics}
        \label{fig:3RGreenGDs}
    \end{subfigure}
     \quad
    \begin{subfigure}[b]{0.3\textwidth}
        \includegraphics[width=\textwidth]{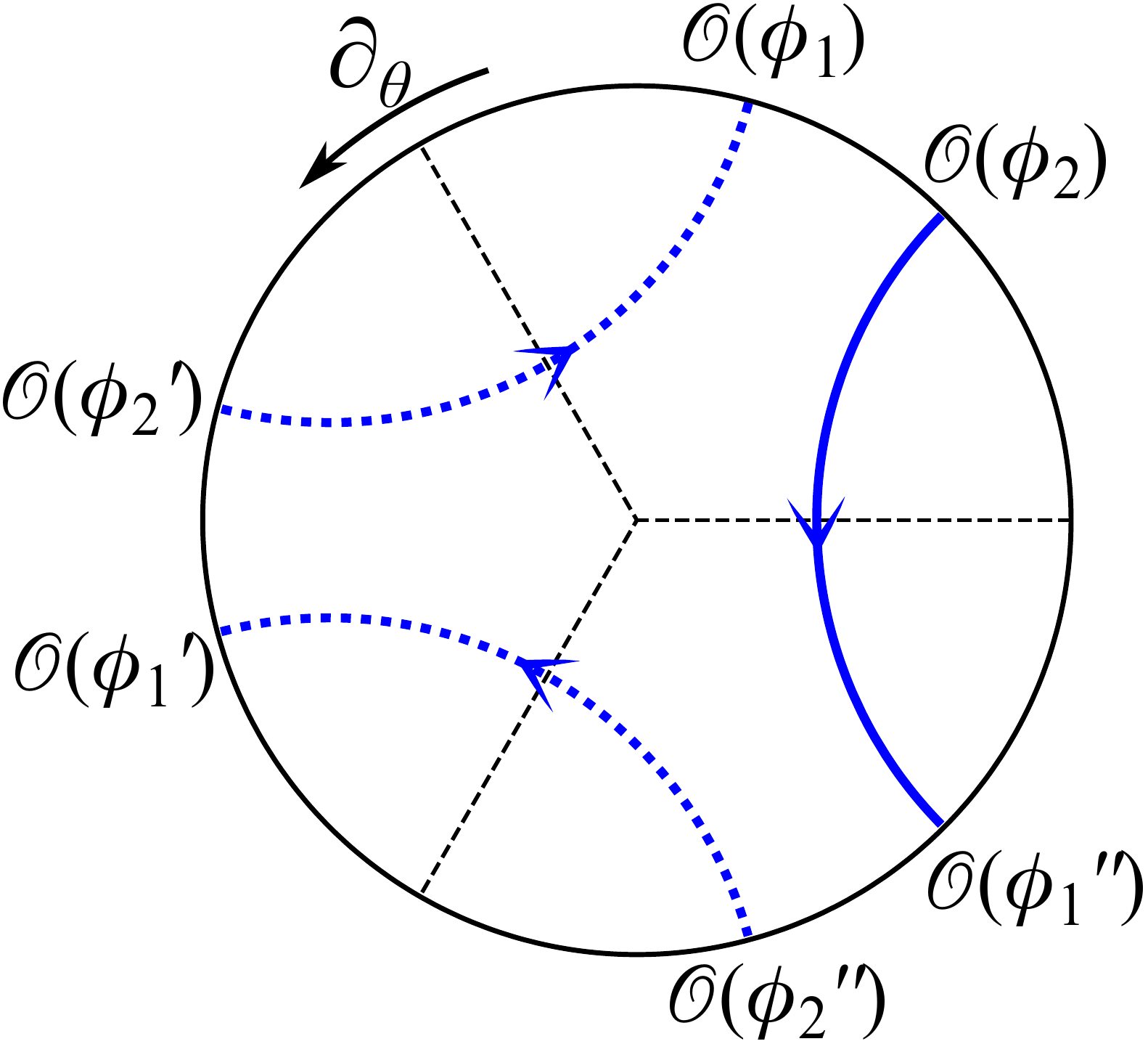}
        \caption{$\mathcal{B}_{k,m=2}$, $n=1$ geodesics}
        \label{fig:3BlueGDs}
    \end{subfigure}
          \caption{(a)-(c) The contributions to the base OPE from symmetrized pairs of operators at fixed angular separation in the covering CFT are encapsulated in the $\mathcal{B}_{k,m}$ blocks. The corresponding oriented bulk geodesics are displayed to show the pairings. Note that only two operators are inserted on the boundary at a time, but all image locations are displayed here for comparison.}\label{fig:3GDs}
\end{figure}

The form of eq. \eqref{eqphiteexpansion} suggests the definition of a more fine-grained OPE block which is symmetrized on the cover,
\begin{equation}\label{eqbkn}
{\mathcal{B}}_{k,m}\left(t,\al,\te\right)=\frac{1}{N}\left|2-2\cos(2\al)\right|^{-\De}\sum_{b=0}^{N-1} \exp{\left(i\frac{2\pi b}{N}\frac{\pa}{\pa  \te}\right)}{\mathcal{B}}_{k}\left(t, \al,\te\right),
\end{equation}
where $\al$ takes on a fixed value $\al_m$ within each term of this block. We emphasize that since this ``partial" OPE block is $\mathbb{Z}_N$ symmetrized it is a valid observable on the base theory. The OPE of the base theory is then put in the suggestive form (cf. \eqref{eqsym})
\begin{align}\label{eqbk}
\begin{aligned}
\tilde{\mathcal{O}}_i(t, \tilde\phi_1) \tilde{\mathcal{O}}_j(t, \tilde\phi_2)  =& \sum_{k} C_{ijk} \frac{1}{N}\sum_{m=0}^{N-1} \exp{\left(i\frac{2\pi m}{N}\frac{\pa}{\pa \phi_1}\right)}  \mathcal{B}_{k,m}(t,\al_m,\te).
\end{aligned}
\end{align}
The partial OPE blocks ${\mathcal{B}}_{k,m}\left(t, \al_m,\te\right)$ encapsulate the contribution to the base OPE from ordered pairs of cover operators at a common distance $\al_m=\al+m\pi/N$, and $\phi_1>\phi_2$ as in figure \ref{fig:3GDs}.

The base OPE blocks $\tilde {\mathcal{B}}_{k}$ in the decomposition \eqref{equsualOPE} group the contributions to the OPE from the conformal family of the primary $\tilde \op_k$. In rearranging the sums to get \eqref{eqphiteexpansion} we lose this interpretation for the partial OPE blocks  $\mathcal{B}_{k,m}(t,\al_m,\te)$. It is not immediately clear what CFT operator contributions these blocks group together. However, we will find that the partial OPE blocks have a clear interpretation in the bulk; they organize the contributions to the base OPE from bulk geodesics of fixed winding numbers.

For each block $ {\mathcal{B}}_{k,m}(t, \al_m, \te)$, the coordinate $ \te$ is in the domain $[0,2\pi/N]$ since $\te=\tilde\te$ was set by convention. We can always choose $ \te$ in this fundamental domain, even though its full domain on the covering space is $[0,2\pi]$, because the symmetry generators in eq. \eqref{eqbkn} permute through all the images of $ \te$ symmetrically. The choice of fundamental domain for this coordinate is the same as the choice for a fundamental domain of kinematic space made in  eq. \eqref{eqdiffentKS} and section \ref{secbulkKS}. 

Importantly, in a single ${\mathcal{B}}_{k,m}(t, \al_m, \te)$ block the coordinate $\al_m$ is restricted to a domain of size $\pi/2N$. To see this, consider the $m=0$ block where the image points have the smallest separation $\al$ and are connected by a geodesic of winding number $n=0$ through the bulk. Fix $\phi_2$ and allow $\phi_1$ to take on different values. Keeping $n=0$ and $\phi_1>\phi_2$ requires $\phi_1$ to stay in the domain $(\phi_2,\phi_2+\pi/N)$. Over this domain $\al_{(m=0)}\in(0,\pi/2N)$ so the $ {\mathcal{B}}_{k,0} $ block corresponds to $\al$ in this range. Increasing $m\to1$ moves the $\phi_1$ insertion to its next image at $\phi_1+2\pi/N$, so the $ {\mathcal{B}}_{k,1} $ block has $\al_{(m=1)}\in(\pi/N,3\pi/2N)$ and corresponds to geodesics of winding number $n=2$. The relationship between $m$ and $n$ is piecewise linear, and differs for even or odd integer $N$. For odd $N$, 
\begin{equation}\label{phi1n-odd}
N \ {\rm{odd:}}\quad 
\begin{array}{| c || c | c | c | c | c | c | c | c |c | c | c |}\hline
m & 0 & 1 & 2 & \ldots & \left\lfloor N/2 \right\rfloor{-}1 &
\left\lfloor N/2 \right\rfloor & 
\left\lceil N/2 \right\rceil & \left\lceil N/2 \right\rceil{+}1 &
\ldots & N{-}2 & N{-}1  \cr\hline
n & 0 & 2 & 4 & \ldots & N{-}3 & N{-1} & N{-}2 & N{-}4 & \ldots & 3 & 1  \cr\hline
\end{array} \,,
\end{equation}
while for even $N$,
\begin{equation}\label{phi1n-even}
N\ {\rm{even:}}\quad
\begin{array}{| c || c | c | c | c | c | c | c | c | c | c |}\hline
m & 0 & 1 & 2 & \ldots & N/2{-}1 & N/2 & 
N/2+1 & \ldots & N{-}2 & N{-}1 \cr\hline
n & 0 & 2 & 4 & \ldots & N{-}2 & N{-}1 & N{-}3 & \ldots & 3 & 1 \cr\hline
\end{array} \,.
\end{equation} 
All values of the winding number $n$ are reached by the $N$ applications of the $\pa/\pa \phi_1$ generator for both odd and even $N$.
In summary, with our conventions each partial OPE block ${\mathcal{B}}_{k,m}(t,\al_m,\te)$ lives in a restricted domain $ \te\in(0,2\pi/N)$ and $\al_m\in (m\pi/N,m\pi/N+\pi/2N)$.

\subsubsection{Partial OPE block Casimir equations}

It was noted in eq. \eqref{eqbilocalcas} that an OPE block satisfies a Casimir equation with the same eigenvalue as the quasiprimary $ \op_k$ it is built from. Since the Casimir operator commutes with all elements of the global conformal group, the $ {\mathcal{B}}_{k,m}$ blocks satisfy the same Casimir equation as the $ {\mathcal{B}}_{k}$ blocks from which they are built \eqref{eqbkn}, with the same eigenvalue, 
\begin{align}
\begin{aligned}
{[\mathcal{C}_2,{\mathcal{B}}_k\left(t,\al, \te\right)]}&= C_k{ \mathcal{B}}_k,\\
\implies [\mathcal{C}_2,{\mathcal{B}}_{k,m}]&=\frac{1}{N}\left|2-2\cos(2\al)\right|^{-\De}\sum_{b=0}^{N-1} \exp{\left(i\frac{2\pi b}{N}\frac{\pa}{\pa  \te}\right)}[\mathcal{C}_2,{\mathcal{B}}_{k}]=C_k {\mathcal{B}}_{k,m}.
\end{aligned}
\end{align}
The differential representation of $\mathcal{C}_2$ must be adapted for the ${\mathcal{B}}_{k,m}$ blocks compared to the $\tilde{\mathcal{B}}_k$ blocks because the conformal generators of the base and cover theory are not the same. 

\begin{figure}
      \centering
        \includegraphics[width=.475\textwidth]{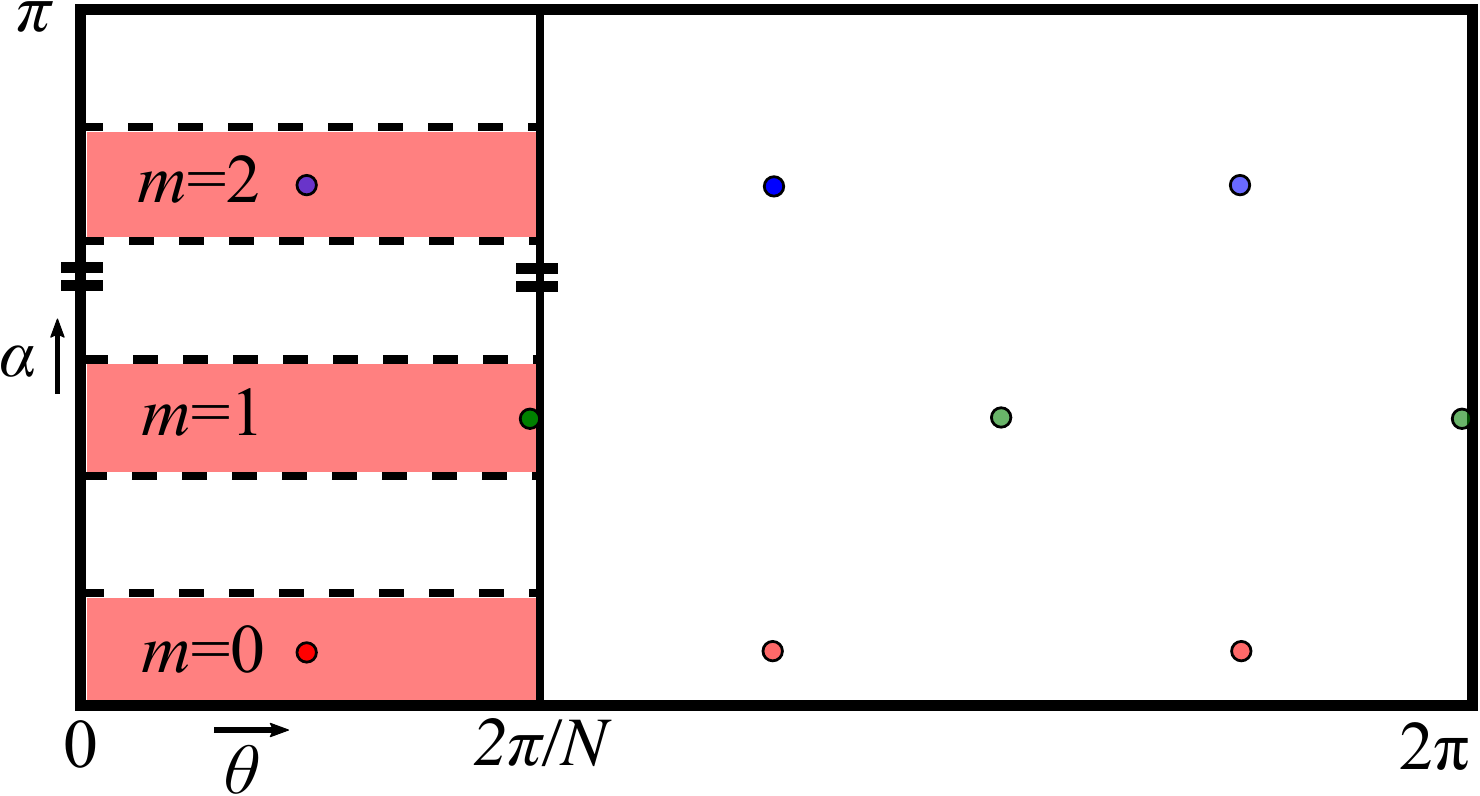}
        \caption{Individual $\mathcal{B}_{k,m}$ blocks give rise to one of the shaded regions of kinematic space. The corresponding geodesics from figure \ref{fig:3GDs} are shown as points. The gaps are filled out by including contributions from the orientation reversed blocks with $\phi_1<\phi_2$. These correspond to the orientation reversed versions of the geodesics in figure \ref{fig:3GDs}.}
        \label{fig:KSCDverticalblocks}
\end{figure}

While the conical defect is dual to an excited state of the base CFT, the covering CFT is in its ground state \cite{Balasubramanian2015}. For this reason, the differential form of the Casimir operator acting on the ${\mathcal{B}}_{k,m}$ blocks is given by eq. \eqref{eqbilocalcas2} using a representation such as in eq. \eqref{eqcylgen}. The only difference that appears in the calculation leading to the Laplacian on kinematic space, eq. \eqref{eqequaltimelaplacian}, is the restricted coordinate domain of ${\mathcal{B}}_{k,m}(\al_m,\te)$: $ \te\in(0,2\pi/N)$ and $\al_m\in (m\pi/N,m\pi/N+\pi/2N)$. Thus the Casimir equation for the ${\mathcal{B}}_{k,m}$ blocks is
\begin{equation}\label{eqcasequationtildebkn}
{[\mathcal{C}_2,{\mathcal{B}}_{k,m}\left(t,\al_m, \te\right)]}=-4\sin^2(\al_m)\left(-\frac{\pa^2}{\pa \al_m^2}+\frac{\pa^2}{\pa\te^2}\right){\mathcal{B}}_{k,m}\left(t,\al_m,\te\right)=C_k{\mathcal{B}}_{k,m}\left(t, \al_m, \te\right),
\end{equation}
which suggests the metric for the kinematic space of the single ${\mathcal{B}}_{k,m}$ block is
\begin{equation}\label{eqksmetricn}
ds_m^2=\frac{1}{\sin^2\al_m}(-d{\al}_m^2+d\te^2).
\end{equation}
This is a subregion of dS$_2$ with the restricted coordinate range as indicated above. Each of the $N$ ${\mathcal{B}}_{k,m}$ blocks gives rise to a region of kinematic space in the same vertical strip of width $\te\in[0,2\pi/N]$ but with differing ranges of $\al$, as depicted in figure \ref{fig:KSCDverticalblocks}. The union of these $N$ regions cover half of the vertical strip, but are not all connected because of how the winding number jumps as one insertion point is permuted through its images, recall tables \eqref{phi1n-odd} and \eqref{phi1n-even}. The indicated half of the vertical strip was obtained by taking $\phi_1>\phi_2$ for the $m=0$ block and acting with $\pa/\pa\phi_1$ generators. By starting with $\phi_1<\phi_2$ for the $m=0$ block and following the same construction with $\pa/\pa\phi_2$ in the place of $ \pa/\pa\phi_1$, one fills out the remaining regions of kinematic space. This is made more clear with a view of the bulk picture in figure \ref{fig:3GDs} where interchanging the roles of $\phi_1$ and $\phi_2$ reverses the orientation of the connecting geodesics.

The base OPE in eq. \eqref{eqbk} receives contributions from each ${\mathcal{B}}_{k,m}$ with both $\phi_1>\phi_2$ and $\phi_1<\phi_2$. Taking the union of the regions identified from each ${\mathcal{B}}_{k,m}$ shows that the kinematic space for the excited states dual to a timeslice of AdS$_3/\mathbb{Z}_N$ can be identified as de Sitter with an identified angular coordinate $ \te=\te+2\pi/N$. In other words, the kinematic space of a static conical defect is a quotient of the kinematic space for pure AdS$_3$, as anticipated in \cite{Czech2015,Czech2016,Asplund2016}. This is the same kinematic space geometry, up to the choice of fundamental region, that was determined from the differential entropy prescription of eq. \eqref{eqdiffentKS}, and the analysis of boundary anchored geodesics under the $\mathbb{Z}_N$ quotient in section \ref{secbulkKS}. 

Just as kinematic space from the bulk point of view can be divided into regions by the winding number of geodesics as in figure \ref{fig:KSCDVertical}, from the CFT perspective kinematic space is built up from the contributions to the OPE by images of fixed separation $\al_m$. This suggests that there should be a connection between the partial $\mathcal{B}_{k,m}$ OPE blocks and geodesics of a fixed winding number associated to $m$. This connection will be made explicit in the following section.

\section{Discussion}\label{secdis}

In this paper we have shown that the kinematic space for a constant time slice of a static conical defect spacetime is a quotient of the kinematic space for time slices of pure AdS$_3$. This fact was anticipated in \cite{Czech2015c,Czech2016,Asplund2016} since all locally AdS$_3$ spacetimes can be obtained as a quotient of AdS$_3$ itself, with geodesics of AdS$_3$ descending to geodesics of the quotient space. From the bulk our results were derived from the original differential entropy prescription, and by studying how the quotient acts on geodesics. The two approaches led to different subregions of the full dS$_2$ kinematic space for pure AdS$_3$, but it was argued that the subregions were equivalent fundamental domains under the identifications. 

From the CFT point of view kinematic space had previously been defined as the space of ordered pairs of points. For a CFT dual to pure AdS$_3$ there is a one-to-one correspondence between ordered pairs of points and bulk geodesics, making it consistent with the bulk definition. Then, conformal symmetry can be used to derive a unique metric on the space of pairs of points, matching the bulk results. However, the one-to-one correspondence is not a typical feature of locally AdS$_3$ spacetimes. While the possibility of including non-minimal geodesics in the description of kinematic space has been considered previously from the bulk \cite{Czech2015c,Asplund2016,Zhang2017}, there has been no clear generalization of the boundary point of view. In this paper we showed that the metric of the kinematic space for conical defects can be inferred from the Casimir equation of partial OPE blocks. Excited states in a discretely gauged CFT dual to conical defects can be related to the ground state of a covering CFT, and gauge invariant operators in base descend from symmetrized operators in the cover. This allows the base OPE blocks to be broken up into distinct contributions from pairs of image operators on the cover at each possible angular separation. These contributions are encapsulated in partial OPE blocks which were shown to satisfy a wave equation. The Laplacian appearing in the wave equation is that of a subregion of dS$_2$, which allows us to infer the metric of patches of kinematic space. The base OPE is a sum of partial OPE blocks, while the union of patches matches the kinematic space identified by bulk arguments. 

The method of images provides the solution to the lack of a one-to-one correspondence between pairs of points and geodesics in this case. When both the bulk and boundary are lifted to their covering spaces, non-minimal geodesics become minimal geodesics connecting distinct image points. The fact that each partial OPE block corresponds to a specific range of $\al$ on the CFT covering space is very similar to how the $\al$ coordinate on kinematic space arranges geodesics by their winding number. This suggests a holographic interpretation for the partial OPE blocks: the block ${\mathcal{B}}_{k,m}$ represents the contribution to the base OPE from a single class of bulk geodesics with fixed winding number $n$ related to $m$ by tables \eqref{phi1n-odd} or \eqref{phi1n-even}. Thus the partial OPE blocks allow for a more fine-grained understanding of the holographic contributions to the OPE. To confirm this suspicion we now consider the holographic dictionary entry relating OPE blocks and bulk fields integrated over geodesics that was established in \cite{Czech2016}, and find that partial OPE blocks are dual to bulk fields integrated over individual minimal or non-minimal geodesics. 

\subsection{Duality between OPE blocks and geodesic integrals of bulk fields}

In \cite{Czech2016} it was noted that a bulk scalar field integrated over a geodesic of AdS$_3$ satisfies the same differential equation on kinematic space as a scalar OPE block.\footnote{See also \cite{Cunha2016,Guica2016} for an independent development of the connection between geodesic operators and OPE blocks.} By verifying that the two quantities also obeyed the same initial conditions a holographic dictionary entry was established for pure AdS$_3$: \emph{OPE blocks are dual to integrals of bulk local fields along geodesics}. The derivation of this dictionary entry relies heavily on the fact that both pure AdS$_3$ and its kinematic space dS$_2\times$dS$_2$ are homogeneous spaces with the same isometry group. This allowed the authors to derive a kinematic space equation of motion for the integrated field by relating the action of the isometries on the field and on the geodesics. In contrast, the conical defect spacetimes are not homogeneous spaces. The defect traces out a worldline that is not invariant under boosts. Nevertheless, progress can be made on extending the dictionary entry to the conical defect case by working on the covering space. For continuity, we will review the essential points of the derivation of the dictionary entry in pure AdS$_3$. Full details can be found in \cite{Czech2016}.

Consider a massive scalar field $\varphi_{AdS}(x)$ on AdS$_3$ integrated over a boundary anchored geodesic $\Ga$ in a constant time slice of the geometry,
\begin{equation}\label{eqxray}
R[\varphi_{AdS}](\Ga)=\int_\Ga ds \ \varphi_{AdS}(x).
\end{equation}
This ``X-ray" transform of $\varphi_{AdS}(x)$ is naturally viewed as a field on kinematic space because it is a function of geodesics, i.e. points in kinematic space.\footnote{When the integration is performed over an extremal surface in a higher dimensional theory this is known as a Radon transform, used first in a holographic context in \cite{Lin2015}.} 

Let $g$ be an isometry of AdS$_3$. The scalar field is invariant under the isometry but its argument is shifted, $\varphi'_{AdS}(x)=\varphi_{AdS}(g^{-1}\cdot x)$. Integrating the shifted field over a geodesic $\Ga$ is equivalent to integrating the original field over a shifted geodesic $g\cdot \Ga$, noting that all isometries of AdS$_3$ map geodesics into geodesics. In terms of the X-ray transform this is expressed as
\begin{equation}\label{eqshifts}
R[\varphi'_{AdS}](\Ga)=\int_\Ga  ds\ \varphi_{AdS}(g^{-1}\cdot x) =\int_{g\cdot\Ga} ds\ \varphi_{AdS}(x) =R[\varphi_{AdS}](g\cdot \Ga).
\end{equation}
A shift in the argument of $\varphi_{AdS}(x)$ can be compensated by a shift in the argument of $R(\Ga)$.

When $g$ is an element of the isometry group near the identity, the action of $g$ on the field is described by the group generators
\begin{equation}\label{eqisomads}
\varphi'_{AdS}(x)=(1-\om^{AB}L_{AB}^{x})\varphi_{AdS}(x),
\end{equation}
where $L_{AB}^{x}$ is an isometry generator of AdS written with embedding space indices, and $\om^{AB}$ is the antisymmetric matrix parameterizing the isometry. In a similar way, the action of $g$ on the X-ray transform is
\begin{equation}\label{eqisomks}
R[\varphi_{AdS}](g\cdot  \Ga)=(1+\om^{AB}L_{AB}^{\Ga})R[\varphi_{AdS}](\Ga),
\end{equation}
where $L_{AB}^{\Ga}$ is an isometry generator on the kinematic space of geodesics. Applying eqs. \eqref{eqisomads} and \eqref{eqisomks} to eq. \eqref{eqshifts} produces the remarkable intertwining relation of isometry generators
\begin{equation}\label{eqintertwining}
L_{AB}^\Ga R[\varphi_{AdS}](\Ga)=-R[L_{AB}^x \varphi_{AdS}](\Ga).
\end{equation}
Applying the same relation twice produces quadratic Casimirs \eqref{eqc2}, in their respective representations of the isometry group;
\begin{equation}\label{eqintertwinedcasimirs}
\mathcal{C}^{\Ga}_2 R[ \varphi_{AdS}](\Ga)=R[\mathcal{C}^{x}_2\varphi_{AdS}(x)](\Ga).
\end{equation}

The subsequent step of the derivation relies crucially on the properties of homogeneous spaces, as noted in \cite{Czech2016}. For homogeneous spaces the Casimir operator of the isometry group is identified with the scalar Laplacian.\footnote{A homogeneous space can be written as the coset space of its isometry group quotiented by the stabilizer subgroup of a point. The Casimir of the isometry group is the scalar Laplacian for the group's Cartan-Killing metric. The same Laplacian is inherited by the coset space when the Casimir acts on functions that are constant on orbits of the stabilizer group. Note that a point in kinematic space is an AdS geodesic, so the stabiliser subgroup of a geodesic in AdS should be used in the quotient.} This was demonstrated for AdS$_3$ in eq. \eqref{eqcasadslaplacian} and for dS$_2\times$dS$_2$ in eq. \eqref{eqopecas}. On the right side of eq. \eqref{eqintertwinedcasimirs} the Casimir acts on a scalar AdS field so the Casimir is in the bulk scalar representation $-\square_{AdS}$. On the left side the Casimir acts on a function of geodesics so it is in the kinematic space representation $-2(\square_{dS}+\bar{\square}_{dS})$. Using the equation of motion for the bulk field and the definition \eqref{eqxray} leads to an equation of motion for the X-ray transform as a scalar field on kinematic space
\begin{equation}\label{eqxrayeom}
2(\square_{dS}+\bar{\square}_{dS})R[\varphi_{AdS}](\Ga)=R[\square_{AdS} \varphi_{AdS}](\Ga)=R[m^2\varphi_{AdS}](\Ga)=m^2R[\varphi_{AdS}].
\end{equation}
This shows that free bulk scalars integrated over boundary anchored geodesics are free scalar fields propagating on kinematic space. This is the same equation satisfied by the OPE block \eqref{eqOPEblockeom} of a spin zero quasiprimary of dimension given by $-\De(\De-2)=m^2$. The X-ray transform and OPE block also satisfy the same initial conditions on a Cauchy slice which establishes that they are dual quantities \cite{Czech2016}. 

\paragraph{Conical defect case.} Now let us analyze the conical defect case, again restricting to the quotients AdS$_3/\mathbb{Z}_N$. The fact that conical defects are not homogeneous spaces precludes the possibility of running through the previous argument directly. However, it is possible to use the intertwining relations obtained in the pure case and only then perform the $\mathbb{Z}_N$ quotient with an appropriate prescription for the X-ray transform over conical defect fields.

Consider a massive bulk scalar field $\varphi_{CD}$ on AdS$_3/\mathbb{Z}_N$, and similarly $\varphi_{AdS}$ on pure AdS$_3$, each described by the action 
\begin{equation}
S=-\frac{1}{2}\int d^3 x\sqrt{-g}\left((\pa\varphi)^2+m^2 \varphi^2\right) .
\end{equation}
The Klein-Gordon equation in global coordinates, eq. \eqref{eqgads} for pure AdS, and eq. \eqref{eqCDmetric} for the defect, is
\begin{equation}\label{eqKGcompare}
\square\varphi=-\frac{1}{\cosh^2\rho}\pa^2_t\varphi+\frac{1}{\sinh^2\rho}\pa^2_\phi\varphi +\frac{1}{\cosh\rho\sinh{\rho}}\pa_\rho(\cosh\rho\sinh{\rho}\ \pa_\rho\varphi)=m^2\varphi.
\end{equation}
For $\varphi_{AdS}$ the angular coordinate is $\phi\in(0,2\pi)$, while for $\varphi_{CD}$, $\phi$ should be replaced by $\tilde\phi\in(0,2\pi/N)$. 

In either case the solutions are obtained through separation of variables \cite{Arefeva2016}. For example, in the AdS case solutions are $\varphi_{AdS}(t,\rho,\phi)=e^{i\om t}Y_l(\phi)R(\rho)$, with the circular harmonics
\begin{equation}
Y_{l}( \phi)=e^{i l\phi},\quad  Y_{ l}( \phi+2\pi n)= Y_{  l}( \phi), \quad l,n\in \mathbb{Z}.
\end{equation}
Similarly, $\varphi_{CD}(t,\rho,\tilde\phi)=e^{i\om t}\tilde Y_{m}(\tilde\phi)R(\rho)$. The circular harmonic here is $2\pi/N$ periodic,
\begin{equation}
\tilde Y_m(\tilde\phi)=e^{iNm\tilde\phi},\quad \tilde Y_m\left(\tilde \phi+\frac{2\pi n}{N}\right)=\tilde Y_m(\tilde\phi),\quad \quad m,n\in \mathbb{Z}.
\end{equation}
Therefore the $\varphi_{CD}$ modes are a subset of the $\varphi_{AdS}$ modes with $l=Nm$. They are $\mathbb{Z}_N$ symmetric $\varphi_{AdS}$ modes that are solutions of the conical defect Klein-Gordon equation in each $\mathbb{Z}_N$ wedge of the covering space, reflecting the quotient structure of the defect. Appropriately symmetrized modes of AdS will be denoted $\varphi_{\mathbb{Z}}(\phi)$; any $\varphi_{CD}$ can be obtained by restricting some $\varphi_{\mathbb{Z}}$ to a single $\mathbb{Z}_N$ wedge.

\begin{figure}
    \centering
    \begin{subfigure}[b]{0.3\textwidth}
        \includegraphics[width=\textwidth]{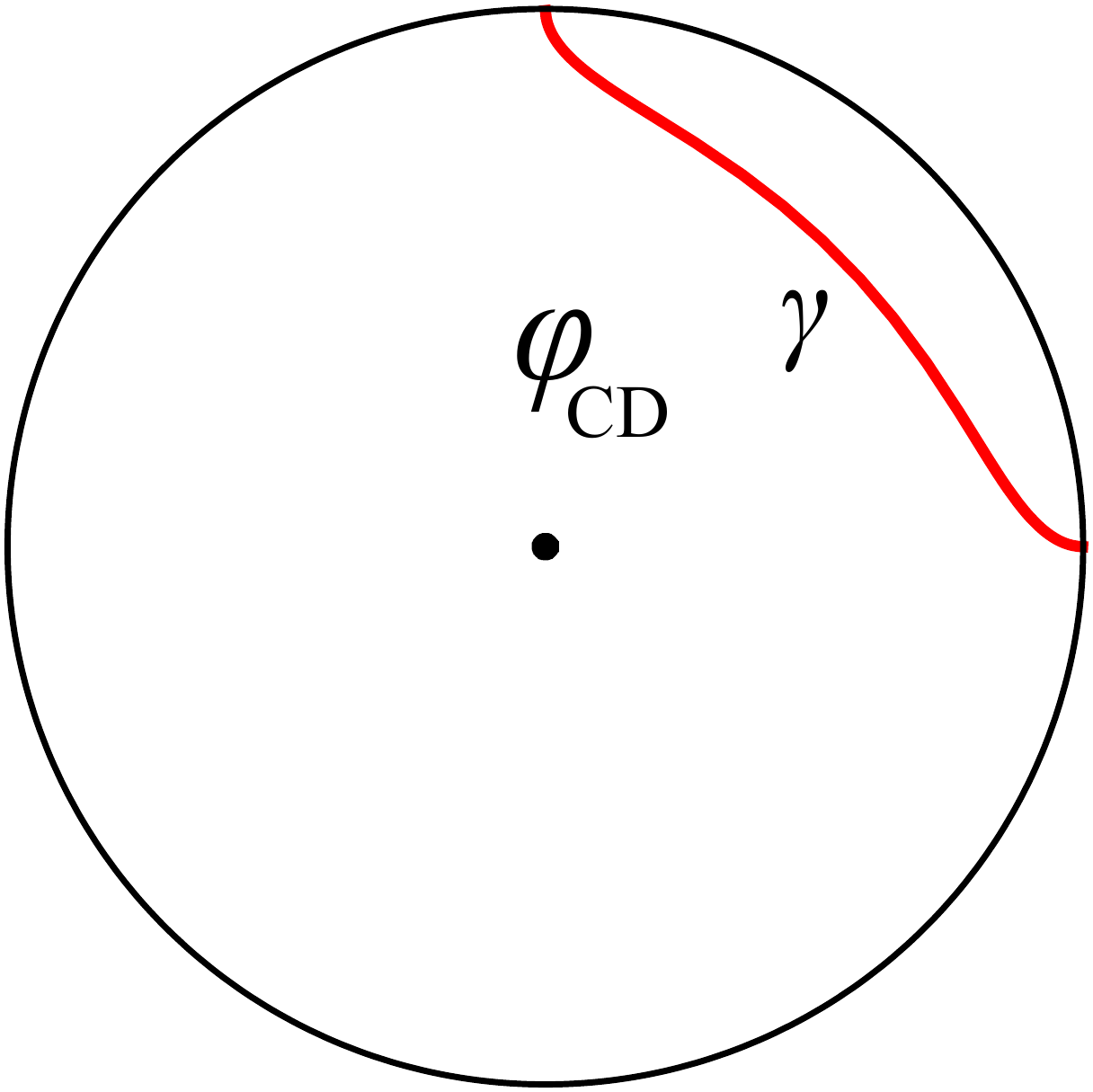}
        \caption{}
        \label{fig:CDGDIntegral}
    \end{subfigure}
     \qquad \qquad 
    \begin{subfigure}[b]{0.3\textwidth}
        \includegraphics[width=\textwidth]{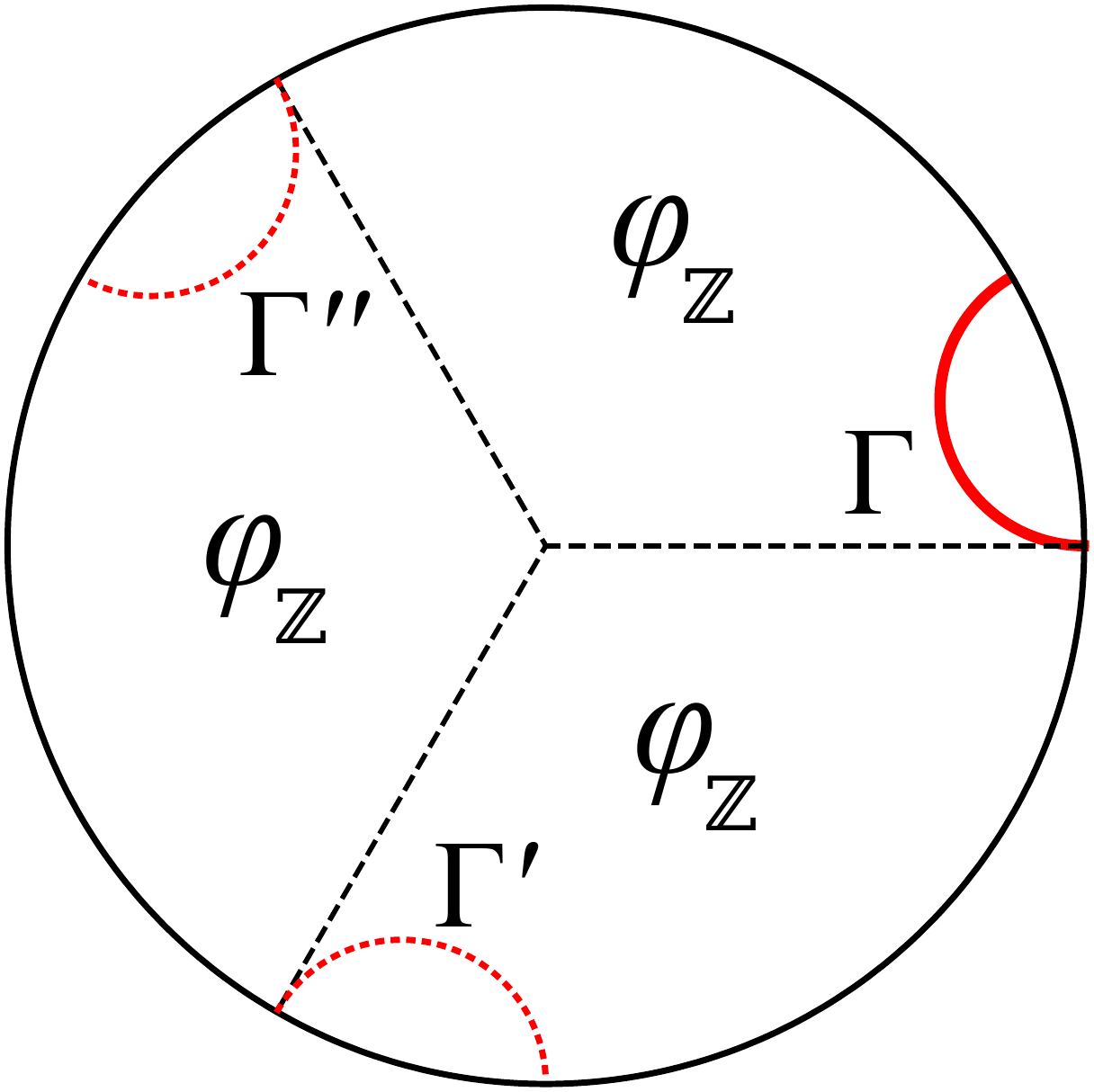}
        \caption{}
        \label{fig:SymGDIntegral}
    \end{subfigure}
    \caption{(a) A conical defect field integrated over a geodesic $\ga$ is the same as (b) a symmetrized AdS$_3$ field integrated over one of the preimages of $\ga$ under the $\mathbb{Z}_N$ quotient. Identifying the edges of any wedge gives the conical defect of (a).}
    \label{fig:6}
\end{figure}

The X-ray transform for the conical defect can then be defined as usual
\begin{equation}
R[\varphi_{CD}](\ga)=\int_\ga ds\ \varphi_{CD}(\tilde x).
\end{equation}
However, this transform acts in a non-homogeneous space and may not share the same invertibility properties as its counterpart eq. \eqref{eqxray}. It is preferable to lift $\varphi_{CD}$ and $\ga$ to the covering space where 
\begin{equation}\label{eqxraycd}
R[\varphi_{CD}](\ga)=\int_\ga ds\ \varphi_{CD}(\tilde x) =\int_\Ga ds \ \varphi_{\mathbb{Z}}( x)=R[\varphi_\mathbb{Z}](\Ga) .
\end{equation}
Instead of integrating $\varphi_{CD}$ over a geodesic $\ga$ in the conical defect spacetime, the corresponding symmetrized AdS field $\varphi_{\mathbb{Z}}$ can be integrated over one of the preimages $\Ga$ of $\ga$ under the $\mathbb{Z}_N$ quotient, see figure \ref{fig:6}. This prescription works for all boundary anchored $\ga$, minimal or non-minimal, since all conical defect geodesics descend from geodesics $\Ga$ on AdS. 

 Note that in going from $\ga$ to $\Ga$, and $\varphi_{CD}$ to $\varphi_{\mathbb{Z}}$ in eq. \eqref{eqxraycd} there is the freedom to choose one of several identical wedges. The choice of wedge will lead to different coordinate values for $\varphi_{\mathbb{Z}}(\phi)$ and $\Ga(\al,\te)$. This is analogous to the ambiguities encountered throughout this paper in choosing a fundamental region. For consistency with the previous choice of a vertical strip of kinematic space, see figure \ref{fig:KSCDVertical}, let $\Ga$ be the preimage of $\ga$ with the smallest centre angle which will always be in the range $\te\in(0,2\pi/N)$. 

By working with the right side of eq. \eqref{eqxraycd}, the properties of homogeneous spaces can be used to find an intertwining relation for the equations of motion. Once again, let $g$ be an infinitesimal isometry of AdS$_3$. The intertwining relation eq. \eqref{eqintertwining} for homogeneous spaces applies as before,
\begin{equation}\label{eqintertwiningcd}
L_{AB}^\Ga R[\varphi_{\mathbb{Z}}](\Ga)=-R[L_{AB}^x \varphi_{\mathbb{Z}}](\Ga),
\end{equation}
and leads to the intertwined Casimirs
\begin{equation}
\mathcal{C}^{\Ga}_2 R[ \varphi_{\mathbb{Z}}](\Ga)=R[\mathcal{C}^{x}_2\varphi_{\mathbb{Z}}(x)](\Ga).
\end{equation}
On the right side the Casimir of AdS isometries becomes the AdS Laplacian which produces the mass eigenvalue. On the left side the Casimir is in the kinematic space representation $-2(\square_{dS}+\bar{\square}_{dS})$ so that
\begin{equation}
2(\square_{dS}+\bar{\square}_{dS}) R[ \varphi_{\mathbb{Z}}](\Ga)=m^2R[\varphi_{\mathbb{Z}}](\Ga).
\end{equation}
On both sides eq. \eqref{eqxraycd} can be used to find the equation of motion for geodesic integrated fields on the conical defect
\begin{equation}\label{eqxraycdeom}
2(\square_{dS/\mathbb{Z}}+\bar{\square}_{dS/\mathbb{Z}}) R[ \varphi_{CD}](\ga)=m^2R[\varphi_{CD}](\ga).
\end{equation} 
The notation $\square_{dS/\mathbb{Z}}$ is to remind that this operator now acts on the subspace of dS$_2$ obtained by restricting to $\te\in(0,2\pi/N)$ with periodic boundary conditions. This is the same behaviour that the symmetrized field exhibits under the quotient, namely $\square_{AdS}\varphi_{AdS}(x)=\square_{CD}\varphi_{CD}(\tilde x)$ within any single wedge.

One might worry that the above argument leading to eq. \eqref{eqintertwiningcd} could break down when $g$ is a boost isometry of AdS under which the conical defect is not invariant. Under the action of a boost, the field $\varphi_{\mathbb{Z}}(g^{-1}\cdot x)$ may no longer be symmetrized around the origin, but the conical defect no longer sits statically at the origin (see \cite{Arefeva2016a} for a relevant discussion). The moving defect is still locally AdS, and can be obtained directly from the covering AdS$_3$ spacetime through an identification along an AdS Killing vector. The identification is no longer a simple angular identification, but shifts time as well as angle. These identifications are given explicitly in \cite{Matschull1999,Balasubramanian1999,Arefeva2015} for example. The moving conical defect solutions can be viewed as global coordinate transformations of the static case, and do not exhibit any different physics compared to stationary ones. On an appropriately boosted timeslice through the moving conical defect spacetime, the transformed field $\varphi_{CD}'(\tilde x)$ can be obtained from $\varphi'_{\mathbb{Z}}( x)$ using the identification that produces the spacetime itself.

The equation of motion for geodesic integrated fields on the conical defect, eq. \eqref{eqxraycdeom}, after taking the equal time limit is the same as the Casimir equation for the base OPE block $\tilde{\mathcal{B}_k}$. The OPE block $\tilde{\mathcal{B}}_k$ represents the contribution to the $\tilde\op_i\tilde\op_j$ OPE  from the conformal family of the quasiprimary $\tilde\op_k$. From the bulk this contribution is obtained by integrating $\varphi$, the dual of $\tilde\op_k$, over \emph{all} geodesics connecting the boundary insertion points of $\tilde\op_i$ and $\tilde\op_j$. This is the well known geodesic approximation which has been used to compute correlation functions \cite{Balasubramanian1999,Arefeva2016a,Goto2017}, and geodesic Witten diagrams \cite{Hijano2016}. Non-minimal geodesics provide a finite number of sub-leading corrections to the minimal geodesic contribution, but can become significant in some regimes.

The connection between bulk and boundary can be made more detailed through the use of kinematic space. Consider the case where $\ga$ is a minimal geodesic. The X-ray transform $R[\varphi_{CD}](\ga_\mathrm{min})$ over a minimal $\ga( \tilde\al,\tilde \te)$, is restricted to $\tilde \al\in(0,\pi/2N)$, $\tilde\te\in(0,2\pi/N)$ with periodicity in the $\tilde \te$ coordinate. The appropriate wave equation \eqref{eqxraycdeom} on the $t=0$ timeslice is
\begin{equation}
4\sin^2 \tilde\al\left(-\frac{\pa^2}{\pa \tilde\al^2}+\frac{\pa^2}{\pa\tilde\te^2}\right)R[ \varphi_{CD}](\ga)=m^2R[\varphi_{CD}](\ga).
\end{equation}
Comparing with eq. \eqref{eqcasequationtildebkn} suggests that the ${\mathcal{B}}_{k,m}$ block with $m=0$ is dual to $R[\varphi_{CD}](\ga_\mathrm{min})$ and represents the contribution to the base OPE from a single class of geodesics, the minimal ones. The duality between ${\mathcal{B}}_{k,0}$ and $R[ \varphi_{CD}](\ga_{\mathrm{min}})$ is established by showing that these quantities satisfy the same initial conditions. The $\tilde\al=0$ Cauchy slice of kinematic space is obtained by taking the coincidence limit of the OPE block, and in the bulk by integrating over a small geodesic that stays near the boundary. These limits are unchanged from the pure AdS case and have been discussed previously \cite{Hijano2016,Czech2016, Boer2016}. In the coincidence limit only the quasiprimary $\op_k$ on the cover, and not its descendants, contributes to the partial OPE block
\begin{equation}
\lim_{\al\to 0}{\mathcal{B}}_{k,0}( \al, \te)=\lim_{\al\to 0} |2 \al|^{\De_k}\op_k(\te),
\end{equation}
while in the conical defect spacetime the behaviour of the dual scalar field near the AdS boundary is given by the extrapolate dictionary
\begin{equation}\label{eqextrapolate}
\lim_{\rho\to \infty}\varphi_{CD}(t=0,\rho,\tilde \phi)=\rho^{-\De_k}\op_k(\tilde \phi),
\end{equation}
so that integrating over a small geodesic localized at $\tilde\phi=\tilde \te$ gives
\begin{equation}
\lim_{\tilde \al\to 0}R[\varphi_{CD}](\ga_{\mathrm{min}}(\tilde\al,\tilde\te))=\lim_{\tilde\al\to 0}\frac{\Ga({\De_k}/{2})^2}{2\Ga(\De_k)}|2\tilde\al|^{\De_k}\op_k(\tilde \te).
\end{equation}
Hence, the initial conditions on kinematic space provide the relative normalization between the dual quantities,

\begin{equation}
R[\varphi_{CD}](\ga_{\mathrm{min}}(\tilde\al,\tilde\te))=\frac{\Ga({\De_k}/{2})^2}{2\Ga(\De_k)}{\mathcal{B}}_{k,0}( \al, \te).
\end{equation}

\begin{figure}
    \centering
    \begin{subfigure}[b]{0.3\textwidth}
        \includegraphics[width=\textwidth]{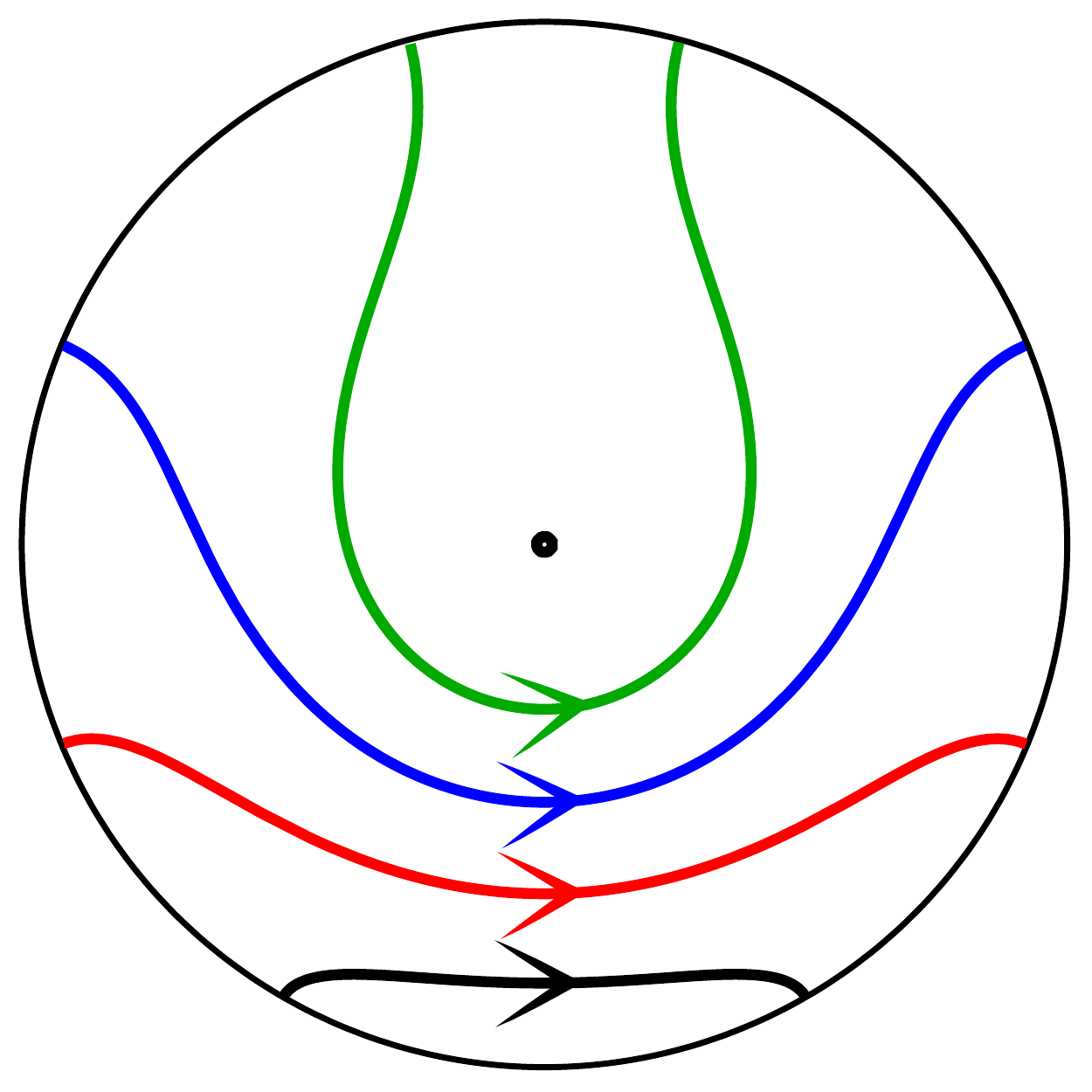}
        \caption{}
        \label{fig:GDContinuity1}
    \end{subfigure}
     \quad 
    \begin{subfigure}[b]{0.3\textwidth}
        \includegraphics[width=\textwidth]{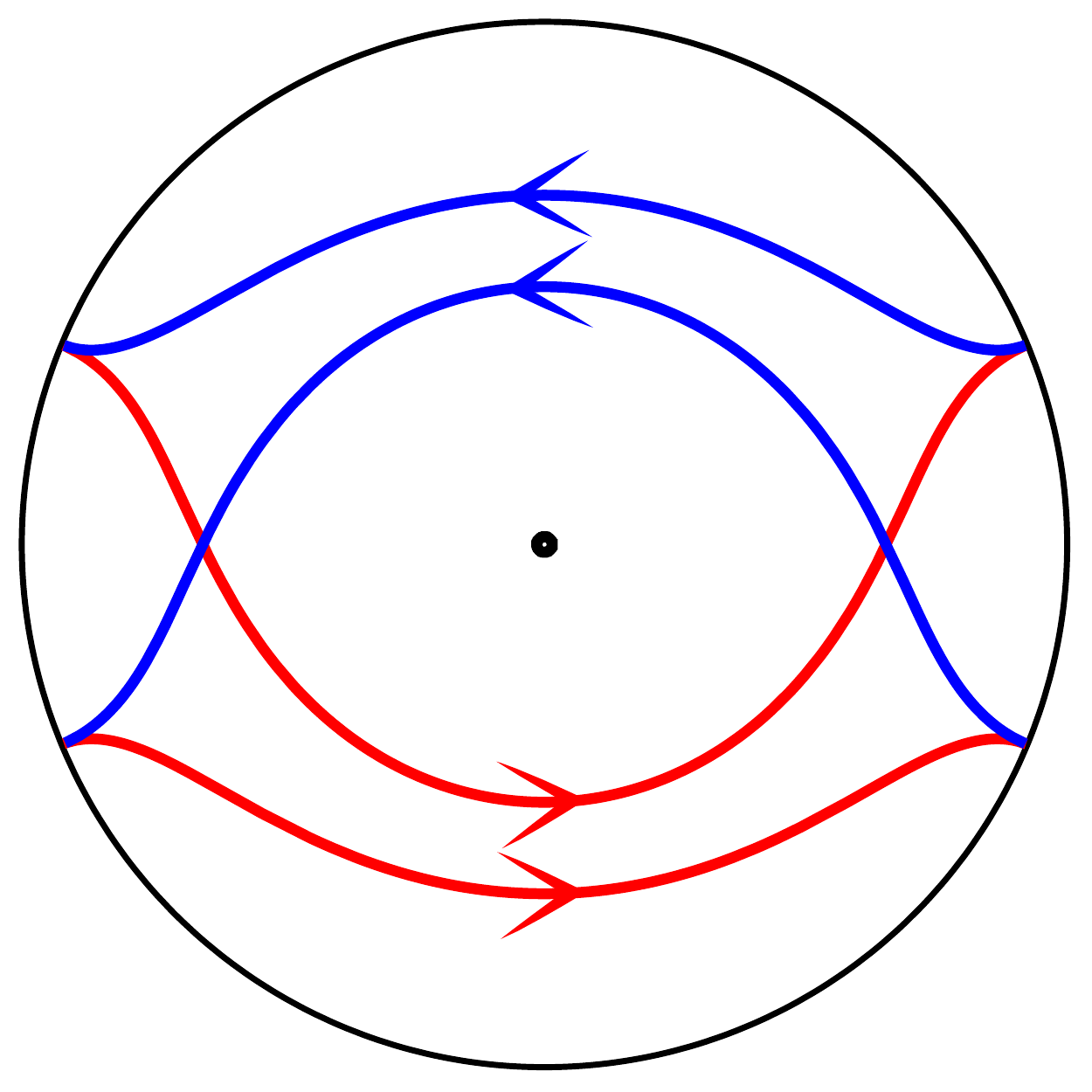}
        \caption{}
        \label{fig:GDContinuity2}
    \end{subfigure}
     \quad 
    \begin{subfigure}[b]{0.3\textwidth}
        \includegraphics[width=\textwidth]{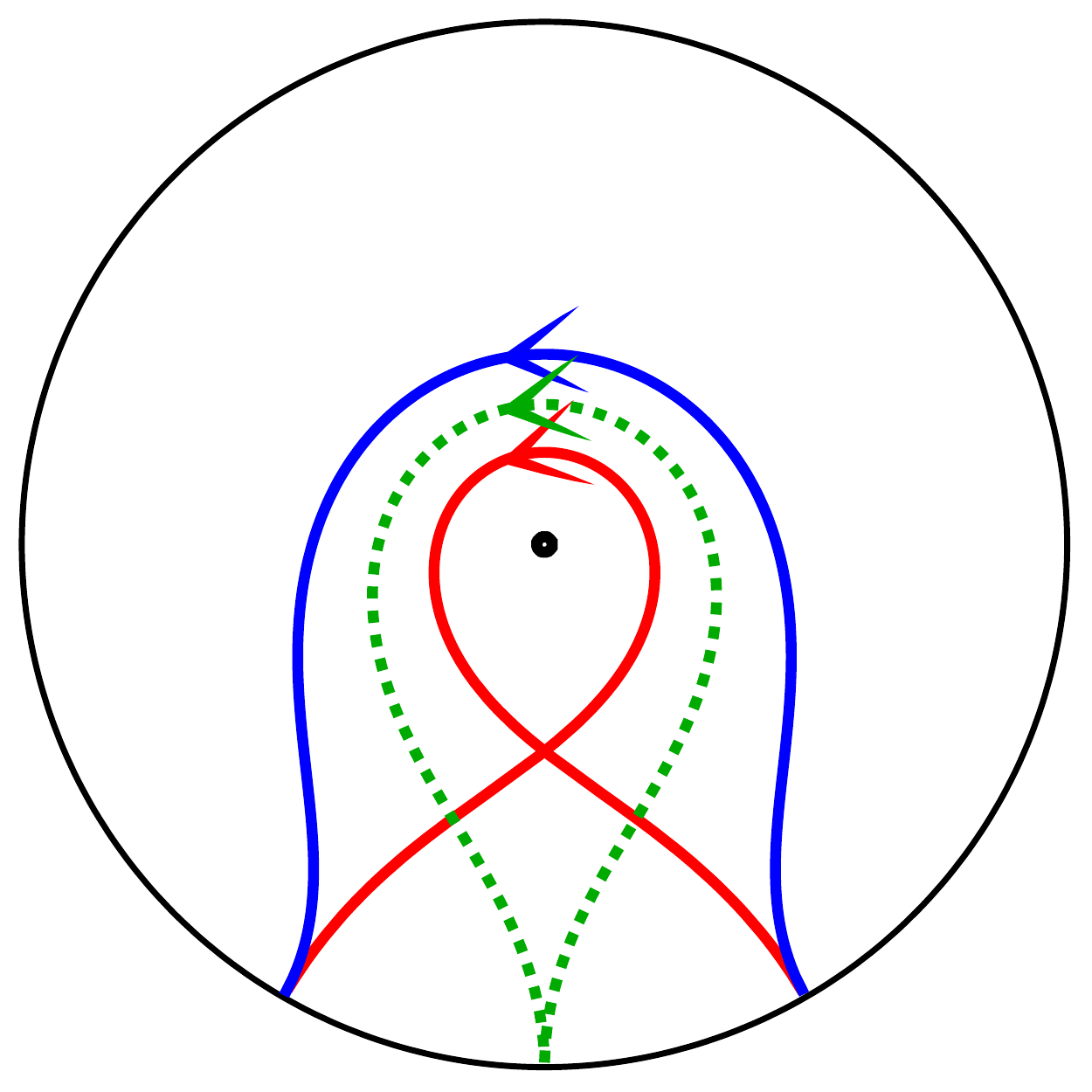}
        \caption{}
        \label{fig:GDContinuity3}
    \end{subfigure}
    \caption{(a) Oriented geodesics away from the defect are continuous in length and shape as their opening angle is increased. (b) Geodesics with the same endpoints but different orientation cannot be smoothly transformed into one another across the defect. (c) As the opening angle of the blue geodesic increases it reaches the dashed geodesic. The red geodesic also reaches the dashed geodesic as its opening angle decreases, showing continuous behaviour even as the winding number jumps.}
    \label{fig:7}
\end{figure}

In general, the ${\mathcal{B}}_{k,m}$ block represents the contribution to the base OPE from the dual field $\varphi_{CD}$ integrated over geodesics with winding number $n$, where $m$ and $n$ are related by table \eqref{phi1n-odd} or \eqref{phi1n-even}. For the non-minimal cases with $n\geq1$, the geodesics do not stay near the boundary, preventing the use of eq. \eqref{eqextrapolate}. However, the transition between winding numbers is smooth. Away from the defect it is clear that there is no discontinuity in the length or shape of \emph{oriented} geodesics as $\tilde \al$ is increased, even as the winding number jumps, see figure \ref{fig:7}. This means the X-ray transform $R[\varphi_{CD}](\ga)$ is a continuous and smooth function of the bulk geodesics on the $\tilde \al<\pi/2$ region of kinematic space. Similarly, the partial OPE blocks ${\mathcal{B}}_{k,m}$ blocks defined in eq. \eqref{eqbkn} are continuous in $\tilde \al$ across transitions in the winding number. This is simply because the OPE behaves smoothly as the operator insertions are moved, and it remains convergent for all separations \cite{Pappadopulo2012}.

\begin{figure}
    \centering
    \begin{subfigure}[b]{0.3\textwidth}
        \includegraphics[width=\textwidth]{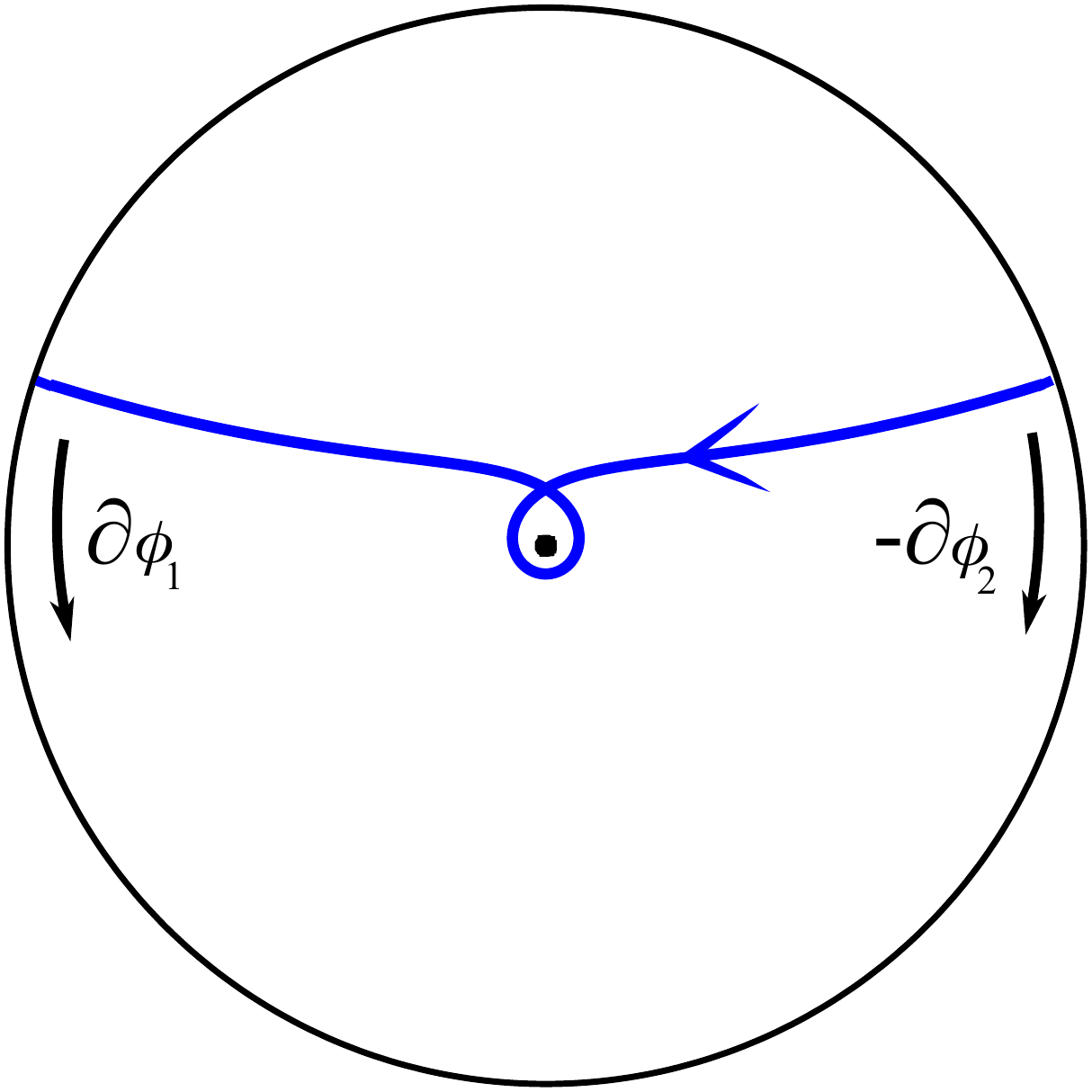}
        \caption{}
        \label{fig:Pinch1}
    \end{subfigure}
     \qquad   \qquad  \qquad
    \begin{subfigure}[b]{0.3\textwidth}
        \includegraphics[width=\textwidth]{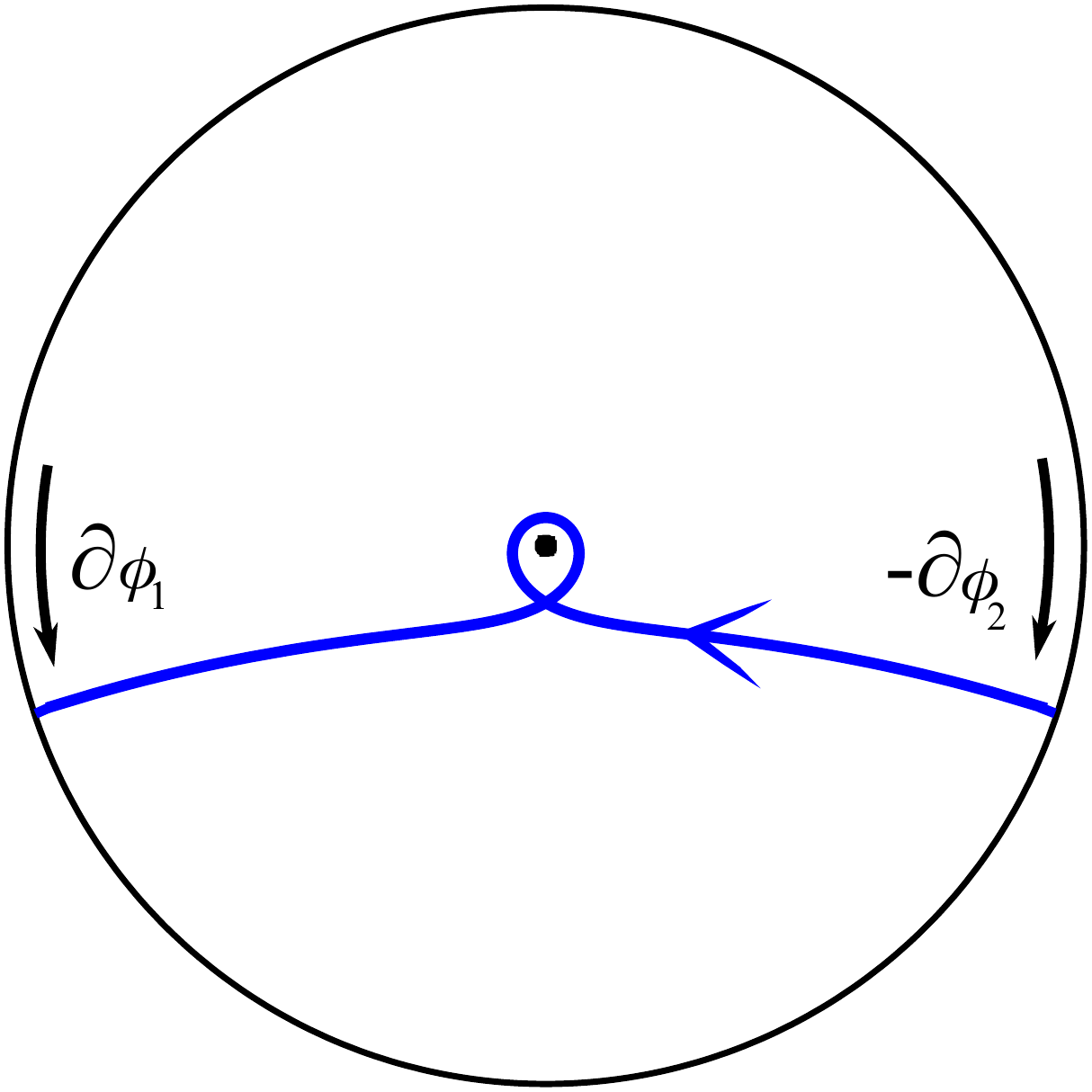}
        \caption{}
        \label{fig:Pinch2}
    \end{subfigure}
    \caption{Geodesics in AdS$_3$ with $\al=\pi/2$ pass through the origin, and descend to geodesics which touch the conical defect. On AdS$_3$, the behaviour of such geodesics is completely smooth as $\al$ is varied, but on the defect spacetime the endpoints may jump as (a) $\al=\pi/2-\ep$ increases to (b) $\al=\pi/2+\ep$ with $\tilde \te$ held constant. This will be the case when $N$ is not an odd integer. Despite this, the length and shape of such geodesics varies smoothly.}
    \label{fig:pinch}
\end{figure}

There is a potential obstacle to the continuity of $R[\varphi_{CD}](\ga)$ at $\tilde \al=\pi/2$ where geodesics touch the defect. Geodesics in AdS$_3$ with $\tilde\al=\pi/2$ pass through the origin and behave smoothly as $\tilde\al$ is varied, but the corresponding geodesics on the defect spacetime must jump as they pinch in on the defect. As depicted in figure \ref{fig:pinch}, geodesics with constant center angle $\tilde\te$ jump as $\tilde\al$ is increased past $\pi/2$ and are not homologous across the jump. Despite this, the length and shape of such geodesics varies smoothly which suggests the X-ray transform of $\varphi_{CD}$ will be smooth as well. That this must be the case is easiest to see by using the lifted X-ray transform \eqref{eqxraycd}. There is no discontinuity whatsoever in the transform of lifted geodesics as $\al$ is increased past $\pi/2$.

One can avoid this obstacle entirely by considering an alternative Cauchy slice on the upper half of kinematic space, namely $\tilde\al=\pi$, which also corresponds to near-boundary geodesics and the coincidence limit for the OPE. By the same argument made for $\tilde\al=0$, the X-ray transform $R[\varphi_{CD}](\ga)$ over a minimal geodesic with $\tilde \al\in((1-1/2N)\pi,\pi)$ obeys the same initial conditions as the corresponding partial OPE block, and both are continuous functions on the $\tilde\al>\pi/2$ half of kinematic space. 

Since the initial conditions in the $\tilde\al\to 0,\pi$ limits match between ${\mathcal{B}}_{k,m}$ and $R[\varphi_{CD}](\ga)$, the equations of motion \eqref{eqcasequationtildebkn} and \eqref{eqxraycdeom} along with continuity in $\tilde\al$ establish the duality between OPE blocks and geodesic operators for static conical defects. The base OPE receives contributions from each bulk geodesic, minimal and non-minimal, connecting the boundary insertion points. Each partial OPE block encapsulates the contribution to the base OPE from the dual bulk field integrated over a single geodesic of fixed winding number.

\subsection{Future directions}

The various approaches to kinematic space used in this paper were adapted to constant time slices of the bulk geometry, equivalently the equal-time limit of the OPE. In each case it was seen that the conical defect kinematic space was a quotient of the pure AdS$_3$ kinematic space using the same quotient that produces the conical defect from AdS$_3$. The full four dimensional geometry of kinematic space describing the time dependent bulk \cite{Czech2016} should also be obtainable using this quotient. There will be a new ambiguity, in addition to the choice of fundamental regions discussed in this paper, from the possibility of rotating in time the faces of AdS$_3$ which are identified, see for example figure 1 from \cite{Arefeva2015}. On neighbouring constant time slices of the AdS$_3$ geometry the wedge representing the conical defect spacetime can have a relative shift in its angular coordinate. The kinematic spaces for subsequent time slices would be vertical strips of dS$_2$ with different ranges of centre angle $\te$. Since the twisted and untwisted identifications of AdS$_3$ produce physically identical conical defect spacetimes, this extra ambiguity can be resolved by making a canonical prescription for an appropriate fundamental region of kinematic space. A complete description of this ambiguity is left for future work.

The CFT results in this paper were derived in the special case dual to AdS$_3/\mathbb{Z}_N$, since there is a particularly simple description of this system in terms of a covering CFT in its vacuum state. It is not surprising that the CFT descriptions of the integer and non-integer cases are significantly different when the holographic consequences are kept in mind. The integer defect spacetimes in the bulk have a mild orbifold singularity that does not obstruct the construction of a consistent string theory on this background \cite{Giusto2013,Balasubramanian2003,Son2001}.\footnote{We thank Oleg Lunin for comments on this point.}

Furthermore, this paper was only concerned with static conical defects. These are part of a more general class of moving defects which are produced either by boosting the static solution, or by taking a quotient of AdS$_3$ along a Killing vector with a timelike component \cite{Matschull1999,Balasubramanian1999,Arefeva2015,Arefeva2016a}. It would be interesting to perform this quotient on the AdS$_3$ kinematic space to obtain the kinematic space of a moving defect. Then, using the relation between OPE blocks and geodesic bulk fields it may be possible to use the method of images to relate back to results on the geodesic approximation for correlation function in those spacetimes.

The partial OPE blocks discussed in this paper reorganize the operator contributions to the base OPE as compared to the traditional OPE blocks. While a clean CFT interpretation of the operator grouping in terms of conformal families is lost, we gain a bulk interpretation in terms of the contributions of geodesics with different winding numbers. It would be enlightening to understand better the CFT operator contributions that are represented by partial OPE blocks. One potential avenue to explore is the superficially similar construction used in \cite{Maloney2017}. Our partial OPE blocks were constructed by first un-gauging a discrete symmetry in going to the covering space description. Gauge invariance is restored by considering symmetrized sums of cover operators under the action of the $\mathbb{Z}_N$ symmetry. The authors of \cite{Maloney2017} studied conformal blocks which give the contribution of a conformal family to a four-point function.  The blocks were approximated by considering only the contribution from light descendants at the cost of modular invariance for the four-point function. Modular invariance was restored by summing over images of the approximate block under the action of modular generators. It may be that these two constructions are related on a deeper level.\footnote{We thank the referee for commenting on this similarity.}

The sum over descendants composing an OPE block evinces that they are non-local operators in the CFT.  As such, OPE blocks ${\mathcal{B}_k}(x_1,x_2)$ have a smeared representation where the quasi-primary $\op_k$ they are built from is integrated over a causal diamond defined by the insertion points $x_1, x_2$ \cite{Czech2016,Boer2016}. It was suggested in \cite{Czech2016} that for conical defects the OPE blocks corresponding to winding geodesics should have a smeared representation over diamonds which wrap all the way around the CFT cylinder (See figure 20 of \cite{Czech2016}). Indeed, our cover OPE blocks have a smeared representation over causal diamonds on the covering CFT cylinder, and so partial OPE blocks can be viewed as symmetrized sums over smeared operators on the cover \eqref{eqbkn}. For the block ${\mathcal{B}}_{k,0}$ representing minimal geodesics, the causal diamonds on the cover are each contained within one $\mathbb{Z}_N$ portion of the cylinder and do not overlap. For blocks representing winding geodesics, the causal diamonds extend over multiple $\mathbb{Z}_N$ portions and can overlap with each other (cf. figure \ref{fig:3RGreenGDs}). Imposing the $\mathbb{Z}_N$ angular identification on any one of these large causal diamonds produces a diamond which wraps around the cylinder of the base CFT and can overlap on itself. It would be interesting to know if the CFT avatar of entwinement \cite{Balasubramanian2015, Balasubramanian2016} can be cast in terms of partial OPE blocks and wrapping diamonds, and how the bulk can be probed in a more fine-grained fashion using these objects.

Other locally AdS$_3$ geometries and their kinematic spaces have been studied from the bulk and using the differential entropy definition \cite{Czech2015b,Czech2015c,Asplund2016,Zhang2017}, but differences in definitions for kinematic space have led to inconsistent results. For instance, the geometries for the kinematic space of the BTZ black holes described in \cite{Zhang2017} include geodesics of both orientations, while \cite{Asplund2016} and \cite{Czech2015b} do not. Furthermore, the authors of \cite{Asplund2016} chose to include only minimal geodesics in their definition of kinematic space, in contrast to the choice we have made here. In this paper we have advocated for defining kinematic space from the CFT in terms of OPE blocks and have isolated the important contributions of non-minimal geodesics. Valuable lessons about the equivalence and applicability of the various kinematic space definitions could be learned by applying this approach to the BTZ case and understanding the kinematic space independently from an OPE block perspective. Work along these directions is in progress.

To date, kinematic space has mainly been used to examine general features of AdS/CFT that can be constrained by conformal symmetry alone. It would be interesting to apply the kinematic space proposal to a particular realization of AdS/CFT. For instance, one could study a chiral primary state in the D1-D5 CFT in the limit where it is dual to a conical defect in the bulk \cite{Lunin2003}.

\acknowledgments

The authors wish to thank James Sully for a fruitful seminar visit, Ian T. Jardine for collaboration in the initial stages of the project, as well as Zaq Carson and Oleg Lunin for useful discussions. The work of JCC and AWP is supported by a Discovery Grant from the Natural Sciences and Engineering Research Council of Canada. JCC is also supported by an Ontario Graduate Scholarship and a Vanier Canada Graduate Scholarship.

\bibliographystyle{JHEP}
\bibliography{KSCD_JHEP}

\end{document}